\documentclass{aa}  
\usepackage{graphicx}
\usepackage{txfonts}
\usepackage{natbib}
\bibpunct{(}{)}{;}{a}{}{,} 

\begin{document}
   \title{Resolving the hot dust around
     HD69830 and $\eta$ Corvi with MIDI and VISIR}

   \author{R. Smith
          \inst{1}
          \and
          M.C. Wyatt \inst{1}
	  \and 
	  C.A. Haniff \inst{2}
          }

   \offprints{R. Smith}

   \institute{Institute of Astronomy, University of Cambridge,
	      Madingley Road, Cambridge, CB3 0AH \\
              \email{rsed@ast.cam.ac.uk}
	 \and 
	     Cavendish Laboratory, University of Cambridge, 
	     JJ Thomson Avenue, Cambridge, CB3 0HE
             }

   \date{Accepted 19th June 2009}

 
  \abstract
   {}
   {Most of the known debris discs exhibit cool dust in regions
    analogous to the Edgeworth-Kuiper Belt. However, a rare
    subset show hot excess from within a few AU, which moreover
    is often inferred to be transient from 
    models for planetesimal belt evolution.
    In this paper we examine 2 such sources to place limits on
    their location to help distinguish between different interpretations
    for their origin.}
   {We use MIDI on the VLTI to observe the debris discs around $\eta$
     Corvi and HD69830 using baseline lengths from 44--130m. New VISIR
     observations of HD69830 at 18.7$\mu$m are also presented. These
     observations are compared with disc models to place limits on
     disc size.}
   {The visibility functions measured with MIDI for both sources
     show significant variation with wavelength across 8--13$\mu$m
     in a manner consistent with the disc flux being well
     resolved, notably with a dip at 10--11.5$\mu$m
     due to the silicate emission feature. 
     The average ratio of visibilities measured between 10--11.5$\mu$m and
     8--9$\mu$m is 0.934$\pm$0.015 for HD69830 and
     0.880$\pm$0.013 for $\eta$
     Corvi over the four baselines for each source, a departure
     of 4 and 9$\sigma$ from that expected if the discs were
     unresolved.
     HD69830 is unresolved by VISIR at 18.7$\mu$m. 
     The combined limits from MIDI and 8m imaging constrain the warm
     dust to lie within 0.05--2.4AU for HD69830 and 0.16--2.98AU for $\eta$ 
     Corvi. }
   {These results represent the first resolution in the mid-infrared
     of dust around main sequence stars.
     The constraints placed on the location of the dust are consistent 
     with radii predicted by SED modelling (1.0AU for HD69830 and 
     1.7AU for $\eta$ Corvi).
     Tentative evidence for a common position angle for the dust
     at 1.7AU with that at 150AU around $\eta$ Corvi, which might be
     expected if the hot dust is fed from the outer disc, 
     demonstrates the potential of this technique for constraining the
     origin of the dust and more generally for the study of
     dust in the terrestrial regions of main sequence stars.}

   
   \maketitle
%

\section{Introduction}

Debris discs are believed to be remnants of planet formation. As well
as providing a unique window on how a system may have formed and
evolved, studies of debris discs can reveal much about the current
status of a system: the location of the dust can indicate possible
configurations of any planetary system around the star \citep[see
  e.g.][]{wyattreview}; and the amount of debris material allows us to
understand what physical conditions may be like for any habitable
planets in the system \citep{greavesbiology}. Debris disc emission is
seen to peak typically longwards of 60$\mu$m implying that the dust is
cool $<$80K and thus predicted to lie at regions many 10s AU from
the star.  In the majority of cases where this emission has been
resolved around Sun-like stars it has been confirmed to lie at $>$40AU 
\citep[see e.g.,][]{holland,greaves,kalas07}. Exoplanet studies on the other
hand typically probe close-in regions for giant planets, as most
planets detected to date have been found through radial velocity
techniques with their inherent detection biases \citep[see
  e.g.][]{butler}.  Debris disc studies are therefore typically
complementary to exoplanet studies.  Furthermore, comparison with
the Solar System's debris discs, the asteroid and Edgeworth-Kuiper
belts,  can help us further understand the likelihood of finding Solar
System analogues in our solar neighbourhood. 

Very few stars exhibit hot dust within $\sim$10 AU, i.e.  in the
region where we expect planets may have formed \citep{ida&lin}.  Four
surveys have searched for hot dust around Sun-like stars by
looking for stars with a 25 $\mu$m flux in excess of photospheric
levels using IRAS \citep{gaidos}, ISO \citep{laureijs} and
Spitzer \citep{hines, bryden}.  All concluded that only $2\pm2$\% of
Sun-like stars have hot dust with infrared luminosities
$f=L_{IR}/L_\star> 10^{-4}$. These hot dust sources provide a unique
opportunity to probe the terrestrial planet region of their systems
\citep[e.g.][]{wyatt02}. \citet{wyattsmith06} identified 7
main-sequence Sun-like (F, G and K-type) sources with hot dust
emission at $\sim$1AU. For these sources the levels of excess emission
observed were compared with the predictions of analytical models for
the collisional evolution of debris discs.  This model predicts that
there is a maximum level of emission that a disc of a given age and
radius can have; discs that are initially more massive, which one may
assume might end up brighter, process their mass through the cascade
more quickly (i.e. more massive discs have shorter collisional
timescales). Although \citet{lohne} found that the analytical model
may underestimate the levels of dust emission produced by debris discs
at ages of $\ga$1Gyr by a factor of 10, they concurred with the
findings of \citet{wyattsmith06} that for 4 of these 7 sources the
level of dust emission is too high to be explained as 
the product of a steady-state collisional cascade in a planetesimal
belt coincident with the dust at $\sim$1AU, and must therefore be the
product of a transient event.  Recent near-infrared interferometric
observations have also revealed 4 systems with hot dust emission at
$\sim$0.1AU from the star (Vega, \citealt{absil}; $\tau$ Ceti,
\citealt{difolco}; $\beta$ Leo and $\zeta$ Lep, \citealt{akeson},
although the hot emission around $\zeta$ Lep must be confirmed). Dust
lifetimes in these regions are very short and these dust populations
are also likely to have been produced in a recent transient event.

A critical diagnostic to understand the origin of this transient
emission is the location of the disc, and in particular whether 
the emission is asymmetric.
In this paper we examine two sources inferred to
have transient emission at $\sim$1AU, HD69830, a K0V-type star with an
age of 2Gyr \citep{beichman06} and $\eta$ Corvi, an F2V-type star with
an age of 1.3Gyr \citep{mallik}. The 
predicted dust locations around both sources is 80--90 mas (from SED
modelling of the IRS spectra of the targets, \citealt{beichman05,
  chen06}), although there remains considerable uncertainty in the
true radial location of the emission.  Thus far
8m imaging of $\eta$ Corvi suggests the dust lies within 3AU of the
star \citep{smithhot}.  Here we present VISIR observations of HD69830
and MIDI observations of both targets.   

This paper is structured as follows.  In section 2 we describe our
interferometric observations and data reduction procedures.  In
section 3 we describe the results of the MIDI observations. In section
4 we present new VISIR imaging of HD69830 showing that the excess
emission at 18.72$\mu$m is not-extended, and discuss the implications
for the maximum extent of the disc. In section 5 we discuss the
results and the implications for the radial location of the debris
discs.  We conclude in section 6.



\begin{table*}
\caption{\label{tab:sources} Characteristics of the science and
  calibration targets.} 
%
%
\begin{center}
%
\begin{tabular}{*{8}{|c}|} \hline \multicolumn{8}{|c|}{Science
  targets} \\ \hline
Source & Spec type & Age & RA & Dec & $F_\ast$ at 10$\mu$m &
$F_{disc}$ at 10$\mu$m & Predicted disc size  \\ 
HD & & Gyr & & & mJy & mJy & mas \\ \hline
69830 & K0V & 2$^a$ & 08 18 23.95 & -21 37 55.8 & 872 & 102 & 80  \\
$\eta$ Corvi & F2V & 1.3$^b$ & 12 32 04.23 & -16 11 45.6 & 1736 & 371
& 90 \\  \hline
\multicolumn{8}{|c|}{Calibrators} \\ \hline
Source & \multicolumn{2}{|c|}{Spectral type} & RA & Dec &
\multicolumn{2}{|c|}{$F_\ast$ at 10 $\mu$m} & Angular size  \\ 
HD & \multicolumn{2}{|c|}{} & & & \multicolumn{2}{|c|}{mJy} & mas
\\ \hline 
61935 & \multicolumn{2}{|c|}{G9III} & 07 41 14.83 & -09 33 04.10 &
\multicolumn{2}{|c|}{9490} & 2.24$\pm$0.01 \\
73840 & \multicolumn{2}{|c|}{K3III} & 08 40 01.47 & -12 28 31.30 &
\multicolumn{2}{|c|}{12312} & 2.40$\pm$0.01 \\
95272 & \multicolumn{2}{|c|}{K1III} & 10 59 46.46 & -18 17 55.56 &
\multicolumn{2}{|c|}{9510} & 2.24$\pm$0.01  \\
116870 & \multicolumn{2}{|c|}{K5III} & 13 26 43.17 & -12 42 27.60 &
\multicolumn{2}{|c|}{10416} & 2.58$\pm$0.01  \\ 
107218 & \multicolumn{2}{|c|}{M4III} & 12 19 42.59 & -19 11 55.97 &
\multicolumn{2}{|c|}{7913} & 2.16$\pm$0.02 \\ \hline
\end{tabular}
%
\end{center}
  For science targets the stellar flux was determined by Kurucz model
  profiles scaled to the 2MASS K band photometry. Total fluxes taken
  from IRS photometry of the targets were then used to determine the
  disc flux, with the predicted
  size being based on SED fitting with a blackbody. For calibration
  targets the angular size is as given by the CalVin tool available at
  http://www.eso.org/instruments/midi/tools. Key: $^a$ Age from
  \citet{beichman06}; $^b$ Age from \citet{mallik}: note that 
  the X-ray luminosity of this star is close to the mean value for the
  Hyades cluster which may suggest a younger age of 600--800Myr
  \citep{stern95}.   
%
\end{table*}

\begin{figure*}
\begin{minipage}{8cm}
\includegraphics[width=8cm]{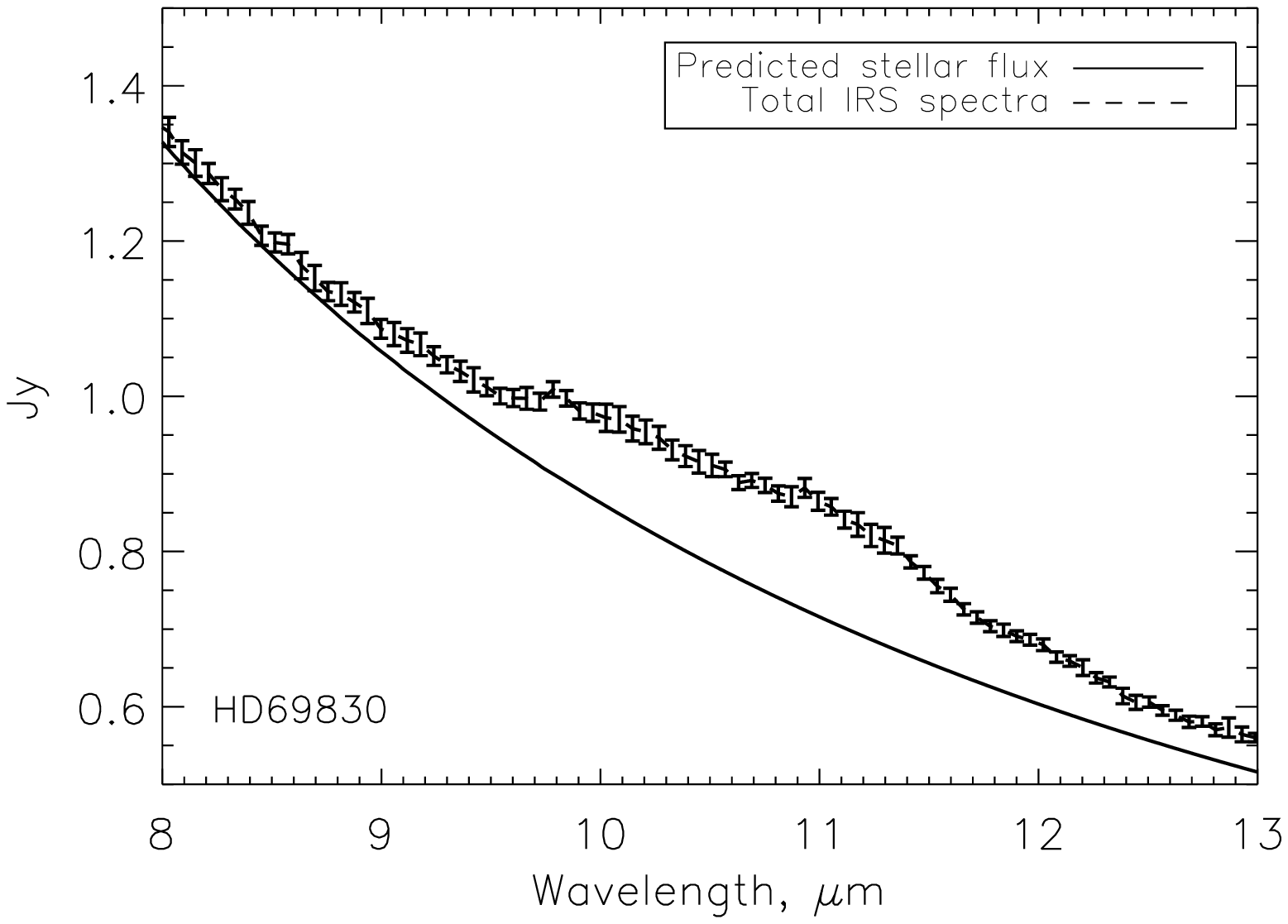}
\end{minipage}
\begin{minipage}{8cm}
\includegraphics[width=8cm]{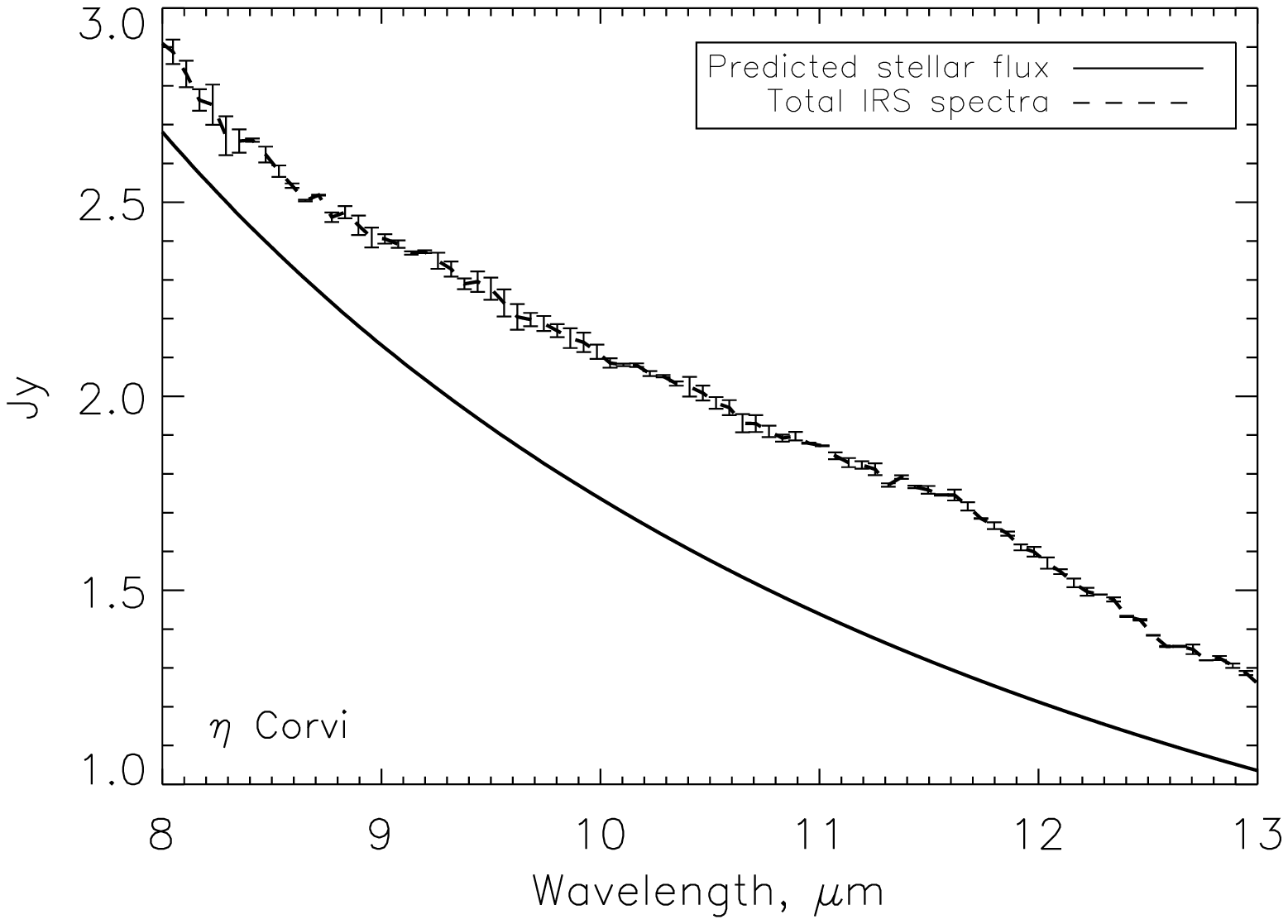}
\end{minipage}
\caption{\label{fig:seds}  The emission spectra of our science targets
  in the MIDI wavelength range.  Left: The Spitzer IRS spectrum of
  HD69830 presented in \citet{beichman05} shows that the excess
  emission has a large silicate feature in the N band. Right: The
  Spitzer IRS spectrum of $\eta$ Corvi also shows a large silicate
  feature at $\sim 11\mu$m.  This spectrum was originally presented in
  \citet{chen06}.  For both sources the emission expected from the
  stellar photosphere is taken from a scaled Kurucz model
  profile. }
\end{figure*}

\begin{table}
\caption{\label{tab:obs} Observations summary}
\begin{center}
\begin{tabular}{*{6}{|c}|}\hline Date and  & Baseline & Pos. Angle &
  Seeing & Source & Target \\ Config. & m & $^\circ$ EoN &
  \arcsec & HD & type \\
  \hline 
05/03/2007 &  98.9 &  40.5 & 0.60 & 61935 & Cal$^a$ \\
UT1-UT3    &  96.4 &  40.9 & 0.55 & 69830 & Sci \\
D1         & 102.1 &  38.4 & 0.95 & 95272 & Cal \\
           & 101.8 &  39.5 & 1.30 & $\eta$ Corvi & Sci \\
           & 102.4 &  38.0 & 1.50 & 116870 & Cal \\ \hline
07/03/2007 &  59.5 & 113.4 & 1.15 & 61935 & Cal \\
UT3-UT4    &  60.4 & 113.7 & 1.30 & 69830 & Sci \\
A          &  59.8 & 114.3 & 1.25 & 73840 & Cal$^b$ \\
           &  58.0 & 105.4 & 0.75 & 116870 & Cal \\
           &  62.4 & 109.0 & 0.80 & $\eta$ Corvi & Sci \\ 
           &  62.1 & 112.5 & 0.65 & 107218 & Cal \\ \hline
08/03/2007 & 130.2 &  63.1 & 0.90 & 61935 & Cal \\
UT1-UT4    & 130.2 &  62.8 & 0.75 & 69830 & Sci \\
C          & 130.1 &  63.3 & 0.75 & 73840 & Cal \\
UT1-UT3    & 100.2 &  31.5 & 1.15 & 116870 & Cal \\
D2         & 102.3 &  38.3 & 1.60 & $\eta$ Corvi & Sci \\ 
           & 102.2 &  36.5 & 0.70 & 116870 & Cal \\ \hline
09/03/2007 &  42.7 &  32.5 & 1.50 & 61935 & Cal \\
UT2-UT3    &  44.4 &  43.8 & 1.55 & 69830 & Sci \\
B          &  44.2 &  34.3 & 1.20 & 73840 & Cal \\
           &  42.2 &  26.8 & 0.65 & 116870 & Cal \\
           &  46.1 &  39.2 & 0.75 & $\eta$ Corvi & Sci \\ 
           &  46.6 &  42.4 & 1.15 & 107218 & Cal \\ \hline
\end{tabular}
\end{center}
  Baselines and position angle (projected on the sky) are
  given as the mean for the observation. \\ $^a$ The first standard
  star observation on run D1 (of HD61935) showed great variation in
  visibility with wavelength (see section 3.4), and so this was
  rejected as a suitable observation for calibration.  \\ $^b$ During run
  A the observation of standard star HD73840 could not be used, as the
  telescope beams showed poor alignment on the readout array. Thus the
  observation of HD69830 was calibrated with HD61935 only, since the
  next usable observation of a standard (HD116870) was taken much
  later in the night after sky conditions had changed significantly. 

\end{table}

\section{MIDI observations and data reduction}

For readers unfamiliar with the basics of optical/infrared
interferometry, an outline of the basics and more detailed information
on the VLTI in particular can be found in \emph{Observation and Data
  Reduction with the VLT Interferometer}, \citep{VLTI}. 

\subsection{Observations}

Interferometric data were obtained on HD69830, $\eta$ Corvi and
suitable standard stars in visitor mode at the VLTI under proposal
078.D-0808 on 5th--9th March 2007.  The expected levels of flux from
our target sources in the N band (see Table \ref{tab:sources} and
Figure \ref{fig:seds}) required the use of the 8m UTs and the HIGH-SENS
observing mode (for details see below). As far as was possible we
tried to observe on both short and long baselines at near
perpendicular position angles so as to constrain the size and geometry of
the sources. In addition, observations of the science targets were
bracketed with observations of standard stars to assess any
variability of the interferometric transfer function.  The science and
standard star observations in this program are summarised in Table
\ref{tab:obs}.

In all cases data were secured following the sequence of steps
outlined below.  A more detailed discussion of this, together with its
rationale can be found in \citet{tristam}, to which the interested
reader is referred for greater detail.

\begin{itemize}

\item 
Initial acquisition, for pointing and beam overlap optimization, was
performed in imaging mode using the N8.7 filter (central wavelength
8.64 $\mu$m, width 1.54 $\mu$m). We used a standard chop throw of
15\arcsec for acquisition and observations of the total source
intensity.

\item 
Subsequently, fringe searching was performed using unfiltered but
dispersed light. A slit width of 0\farcs52 width 
%
was used together with the MIDI prism dispersing element giving a
spectral resolution, $\lambda / \Delta \lambda$, of 30 at $\lambda =
10.6 \mu$m.
%

\item 
After  fringe   centering,  longer  sequences  of   fringe  data  were
secured. For  the first  few frames of  each sequence, the  delay line
positions were altered so as  to be significantly offset from the zero
optical path difference (OPD) position to allow a  determination of thermal
and incoherent backgrounds. Thereafter,  the delay lines were returned
to their nominal locations and typically a few hundred scans of fringe
data were secured. 
These data allowed an ongoing determination of the OPD corrections needed 
to compensate for uncorrected sidereal and atmospheric motions. It this way 
it was ensured that that the fringe packets always remained well centred.

\item 
Finally, at the end of each interferometric data sequence, the same
instrument set-up was used to determine the throughput of the
individual UT optical trains.  For these "intensity calibration"
measurements chopping of the secondary mirror was used to suppress any
thermal background.  The chop frequency and integration time before
readout were varied between the runs to try to maximise the
signal-to-noise on these measurement on a case by case basis.
\end{itemize}

In summary, each full dataset for a single observation of a target
comprised a fringe file (containing the interferometric data from the
combined beams) and 2 ``intensity calibration'' observations (one for
each beam). Similar data were also secured on a standard star
(i.e. of known flux and small diameter) so that flux calibrated
visibilities could extracted. The standards were selected from the
spectro-photometric catalogue available on-line at
http://www.eso.org/instruments/visir/tools and are listed in Table
\ref{tab:sources}.

\subsection{Reduction with the MIA+EWS package}
Data reduction was performed using 
%
the EWS software, available as part of the MIA+EWS package (see,
http://www.strw.leidenuniv.nl/$\sim$nevec/MIDI/index.html and the
manual and details therein).
%

\begin{figure*}
\begin{minipage}{9cm}
\includegraphics[width=9cm]{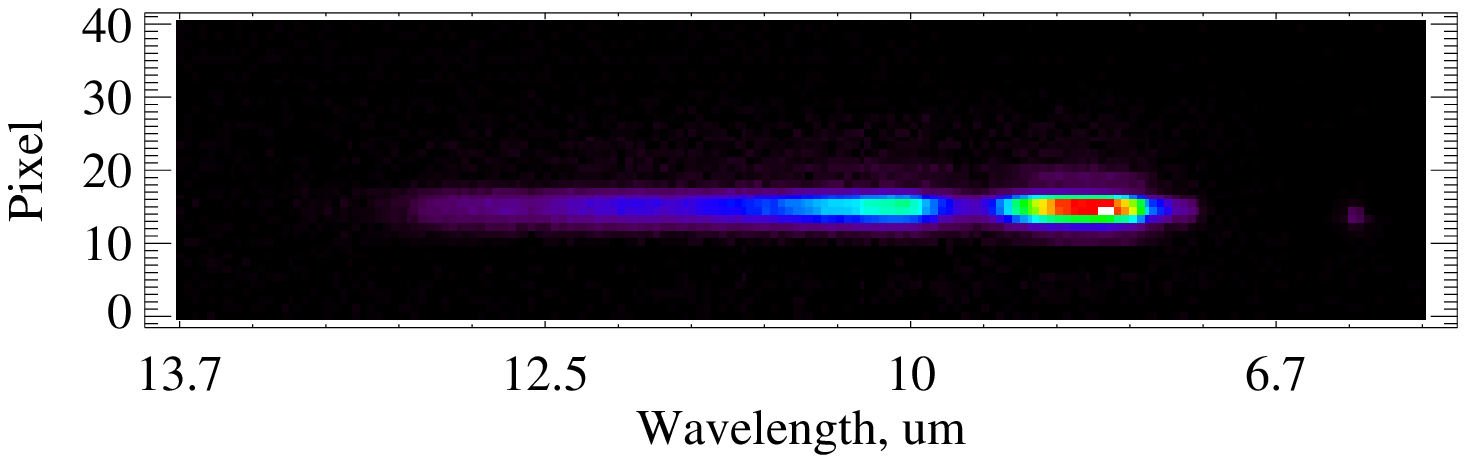}
\end{minipage}
\begin{minipage}{7cm}
\includegraphics[width=7cm]{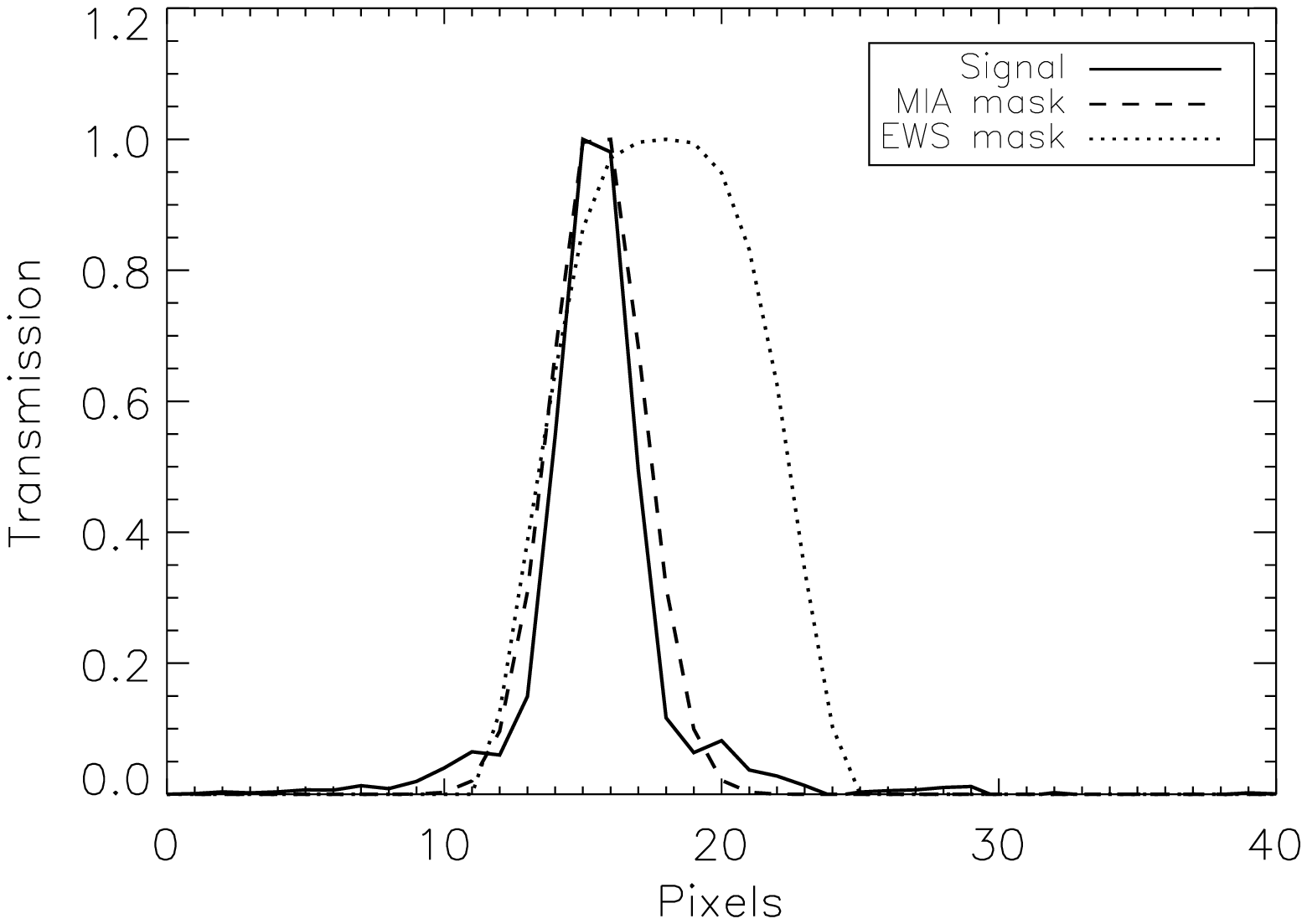}
\end{minipage}
\caption{\label{fig:photframes} Total intensity from MIDI
  observations.  Left: Coadded intensity frames from an observation of
  a standard star in a single optical train (in this case HD61935 run
  D1 beam ``A'' = telescope UT3). The spectral dispersion runs along
  the horizontal direction (increasing wavelength from right to left).
  Right: A cut through one of the spectral channels (at
  8.5$\mu$m). This shows the profile of the total intensity along the
  pixel direction.  Overplotted are the EWS default mask (fixed based
  on observation set-up) and the MIA mask (determined from the
  intensity image itself).  It is evident that the MIA
  mask fits not only the width of the true detected profile but also
  the location of the peak on the array much more accurately than the
  default mask.}
\end{figure*}

The first step in the data reduction involved compressing the fringe
and total intensity frames in the direction perpendicular to the
spectral dispersion to obtain one-dimensional fringe and intensity
spectra.  Prior to compression, each frame was multiplied by a mask to
reduce the impact of noise.  
We found that the default EWS masks provided for
HIGH-SENS mode observing gave a poor fit to the peak of the
detected emission and were much broader than the emission in the
direction perpendicular to the spectral dispersion.  We therefore
used masks determined by a fit to the total intensity frames
as provided by the MIA reduction package instead 
(see Figure \ref{fig:photframes}).
We then used the EWS \emph{group delay analysis} to
align the fringes before vector averaging over time to derive the
\emph{correlated flux} (for details see the EWS user manual). 
The correlated flux $I_{\rm{corr}}$ was then compared to the total
source flux, $I_{\rm{tot}}$, to give the source visibility $V =
I_{\rm{corr}}/I_{\rm{tot}}$.

\subsection{Total Source Intensity}

Total intensity data were obtained with MIDI after the fringe
observations had been secured as described in Sec.~2.1.  The results
showed a great degree of variation between different observations of
the same object, primarily because of background which could not be
perfectly subtracted. We found the background did not vary linearly in
the direction perpendicular to the spectrally-dispersed direction, and
moreover varied with time during the observations, suggesting that sky
(and/or telescope) variations were significant. As the total intensity
measurements of the same science target could vary by $>$50\%, we
chose to use available IRS spectra of our science targets as the total
intensity measurements instead. The IRS spectra (originally presented in
\citealt{chen06} and \citealt{beichman05}) were extracted along a slit
of width 3\farcs7 or 4\farcs7 depending on the spectral
resolution. Our MIDI data were secured with a slit width of 0\farcs52
and so it was necessary to check whether the Spitzer aperture
might have included additional photometric signals not seen by MIDI.

%
For both sources we found there were no background or companion objects in the
%
Spitzer aperture that might be excluded by the MIDI aperture
\citep{smithhot}.  The 3 Neptune mass planets discovered within 1AU
of HD69830 \citep{lovis} are not expected to be brighter in the
mid-infrared than 10$^{-3}F_\star$ \citep{burrows}, and so would
only contribute a discrepancy of at most
$\sim$1mJy at 10$\mu$m to our total intensity measurement. 
%
%
A full analysis of the IRS spectra \citep{lisse07} revealed evidence
for water ices which, due to their lower temperatures, may reside at
larger radial offsets from the star than the bulk of the emission
around HD69830.  However, as pointed out by the authors the local
thermal equilibrium temperature for the region of their best fitting
model --- a disc at $\sim$ 1AU --- is only 245K,
cool enough that should the water ice be isolated from the hotter dust
particles it could be stabilised by evaporative sublimation.  In
addition a wide Kuiper belt-like location is ruled out by the 70$\mu$m
limits on the excess (1$\pm$3mJy, \citealt{beichman05}). The
contribution of the water ice in the 8-13$\mu$m region of the spectrum
is at the level of 1-3\% of the excess emission between 8-11$\mu$m,
rising to 27\% of the excess emission at 13 $\mu$m.  Thus we 
believe the effect of the water ice contribution falling outside the
MIDI beam can only have a significant effect on the measured flux
longwards of 11.5$\mu$m, if at all.  The temperature of the dust
around $\eta$ Corvi is 
predicted by SED fitting to be at around 320K, suggesting a small
radial offset from the star.  The alternative fit to the SED suggested
by \citet{chen06} suggests there may be dust at two temperatures of
360K and 120K, but the 8--13$\mu$m range is dominated in this model by
amorphous olivines at the hotter temperature.  Thus in either the
single temperature fit or the fit proposed by \citet{chen06} the
excess emission in the MIDI wavelength range is expected to be
dominated by grains close to the star.  Therefore for both science
targets it is unlikely that significant emission appears in the IRS
spectra that would be excluded from our MIDI observations.

\subsection{Visibility calibration}

\begin{figure*}
\begin{minipage}{8cm}
\includegraphics[width=8cm]{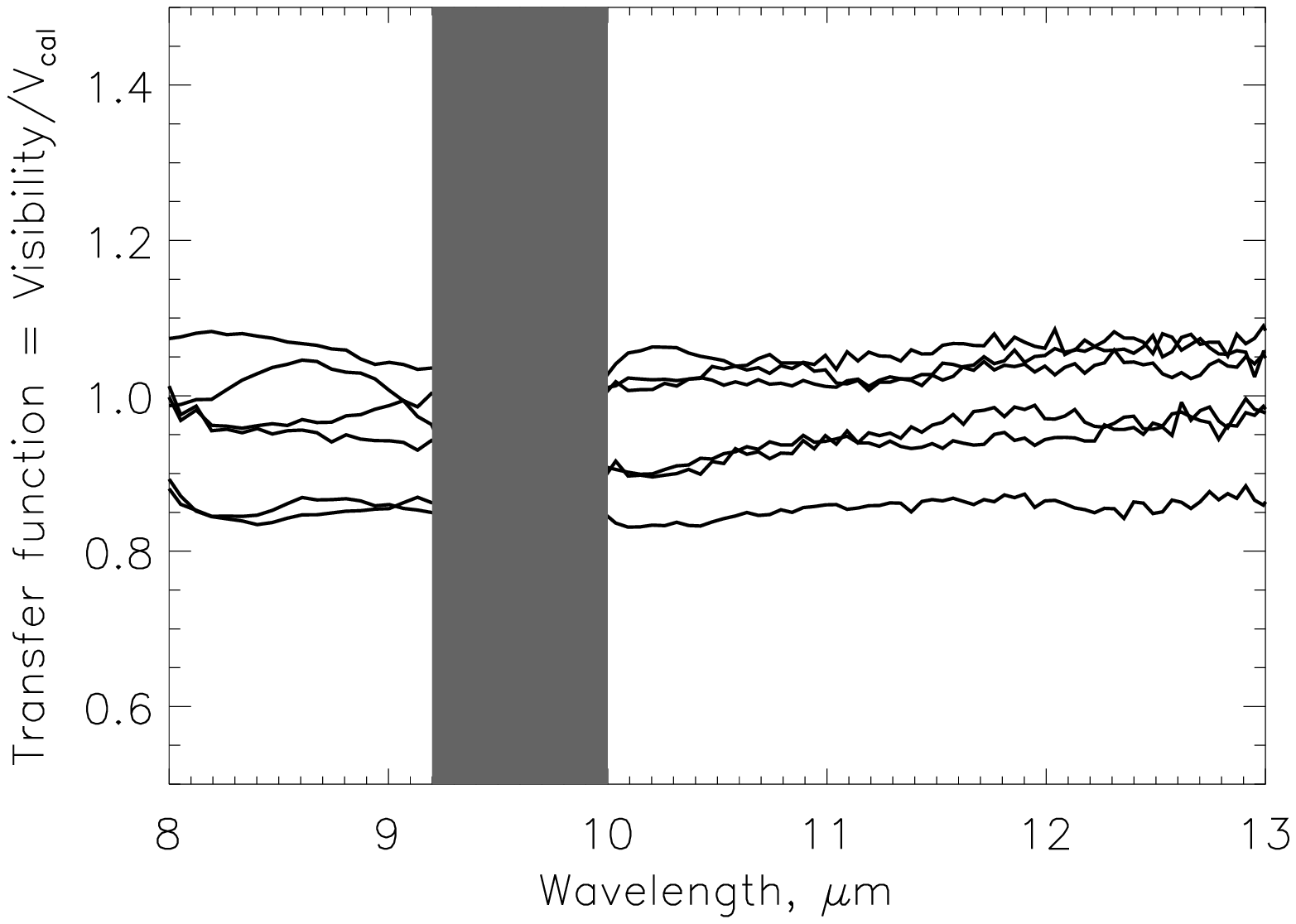}
\end{minipage}
\begin{minipage}{8cm}
\includegraphics[width=8cm]{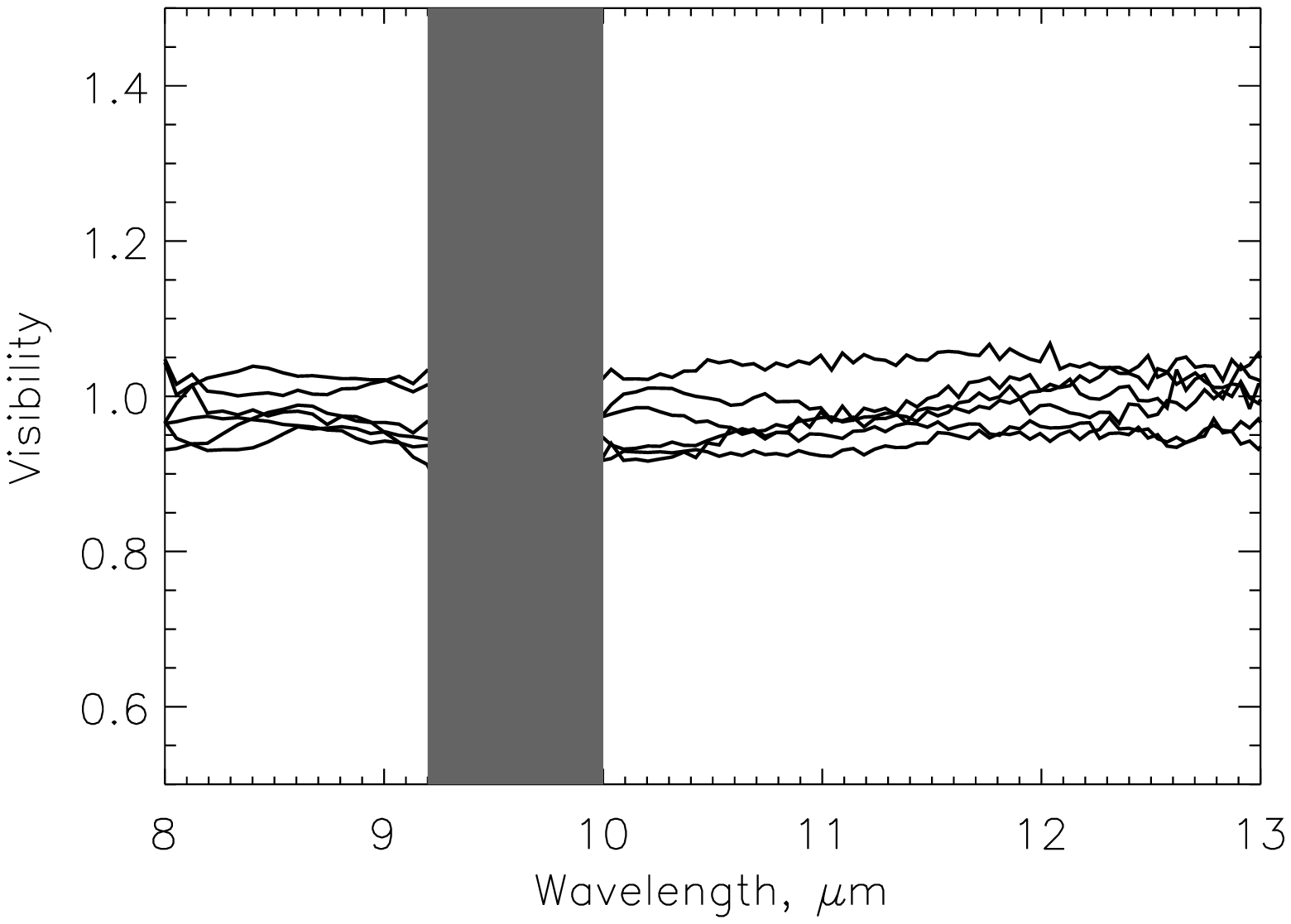}
\end{minipage}
\caption{\label{fig:calcalvis} Plots of visibilities determined for
  standard star targets using other standard stars.  Pairs of standard
  stars observed closely in time were considered as these are the
  pairings used for science observation calibration.  Note that the
  shaded region 9.2--10$\mu$m is prone to strong uncertainty due to
  ozone absorption, and so should be ignored. For each pair the
  'target' and 'calibrator' were chosen randomly.  Left: The
  calibrator-calibrator determined visibilities divided by the
  predicted visibility for the target star.  Perfectly calibrated
  data would give a straight line at unity with no variation with
  wavelength.  We used the weighted mean (over wavelength) for each
  calibrator-calibrator dataset to quantify the quality of the 
  calibration and found a range between 0.86 and
  1.12.  Right: Checking the absolute variation with wavelength for
  calibrator-calibrator visibilities. All visibilities were scaled to
  1 in the range 8--9$\mu$m. The ratio of visibility between
  10--11.5$\mu$m and 8--9$\mu$m was calculated to determine the
  variation of visibility with wavelength.  The maximum deviation from
  unity for this value was 0.047, with some calibrations providing a
  much flatter visibility. }
\end{figure*}

The total source intensities of the standards, which were chosen for
their lack of infrared spectral features as well as their small
angular size, were modelled by Rayleigh-Jeans functions scaled to the
10.5$\mu$m flux listed in the spectro-photometric catalogue (see Table
\ref{tab:sources}).  The visibilities of the standard stars were
calculated assuming these could be modelled as uniform discs with
diameters as listed in Table \ref{tab:sources}.  

The calibration of the correlated flux and determination of
visibilities for each observation were determined as
\begin{equation}\label{eq:calcorr}
 F_{\rm{corr,tar}} = I_{\rm{corr,tar}}/I_{\rm{corr,cal}} \times
F_{\rm{tot,cal}} \times V_{\rm{cal}},
\end{equation} where $I_{\rm{corr,tar}}$ and $I_{\rm{corr,cal}}$ are
the correlated fluxes in ADU (output of data reduction procedure) 
of the target and standard star used as a 
calibrator respectively, $F_{\rm{corr,tar}}$ is the calibrated correlated
flux of the target, $F_{\rm{tot,cal}}$ is the known flux of the
standard star (from Rayleigh-Jeans slope) and $V_{\rm{cal}}$ is the
visibility of the standard star which is dependent on the baseline
length and wavelength.  The visibility of the
target is then 
\begin{equation} \label{eq:calvis}
V = F_{\rm{corr,tar}}/F_{\rm{tot,tar}}, 
\end{equation}
($F_{\rm{tot,tar}}$ again from
Rayleigh-Jeans slope). For the case of the science targets 
 $F_{\rm{tot,tar}}$ was given by the IRS spectrum.  
%
We used the average of the two standards bracketing each science observation
to determine the calibrated visibilities, and difference between the two to
determine our absolute calibration uncertainties.  

We also performed the above calibrations with the standard star
observations as ``dummy targets'' to: 1) determine
the level of accuracy to which we could trust the absolute visibility
measurements; and 2) examine the visibility functions for evidence of
changing visibility with wavelength.  Examples of the transfer
function of standard -- standard
calibration (where the standards were observed close in time and space 
to each other) are given in Figure \ref{fig:calcalvis}. In the left
panel we show the transfer function; the visibility determined for our
``dummy targets'' as a function of the visibility predicted for this
target ($V_{\rm{cal}}$ in  equation \ref{eq:calcorr}).  
%
The absolute levels of this transfer function were found to have a mean and
standard deviation of 0.99 $\pm$0.10 (Figure \ref{fig:calcalvis} left)
after calculating the weighted mean of each visibility over the
8-13$\mu$m range (excluding the 9.2--10$\mu$m region which is
dominated by ozone emission). The weights were determined by splitting
the fringe data into 5 equal length sets and calculating the variance
between these subsets of data at every spectral channel.  This
absolute variation is the same as the 10\% error typically expected in
calibration due to changing conditions between observations
(e.g. \citealt{chesneau}).  The visibility functions were fairly flat,
as shown in the right-hand panel of Figure \ref{fig:calcalvis}.  The
weighted mean of each visibility function between 8-9$\mu$m was used
to scale the visibilities to 1 in the 8--9$\mu$m range for this plot.
The weighted mean for each function between 10--11.5$\mu$m and
8--9$\mu$m were then compared. This ratio was found to be have a mean
and standard deviation of 0.991 $\pm$ 0.037.  Thus the differential
visibility (i.e. the visibility as referenced to that at some fixed
wavelength) is typically known with $\sim$ 3 times better accuracy
than the absolute visibilities.


\begin{figure*}
\begin{minipage}{8cm}
\includegraphics[width=8cm]{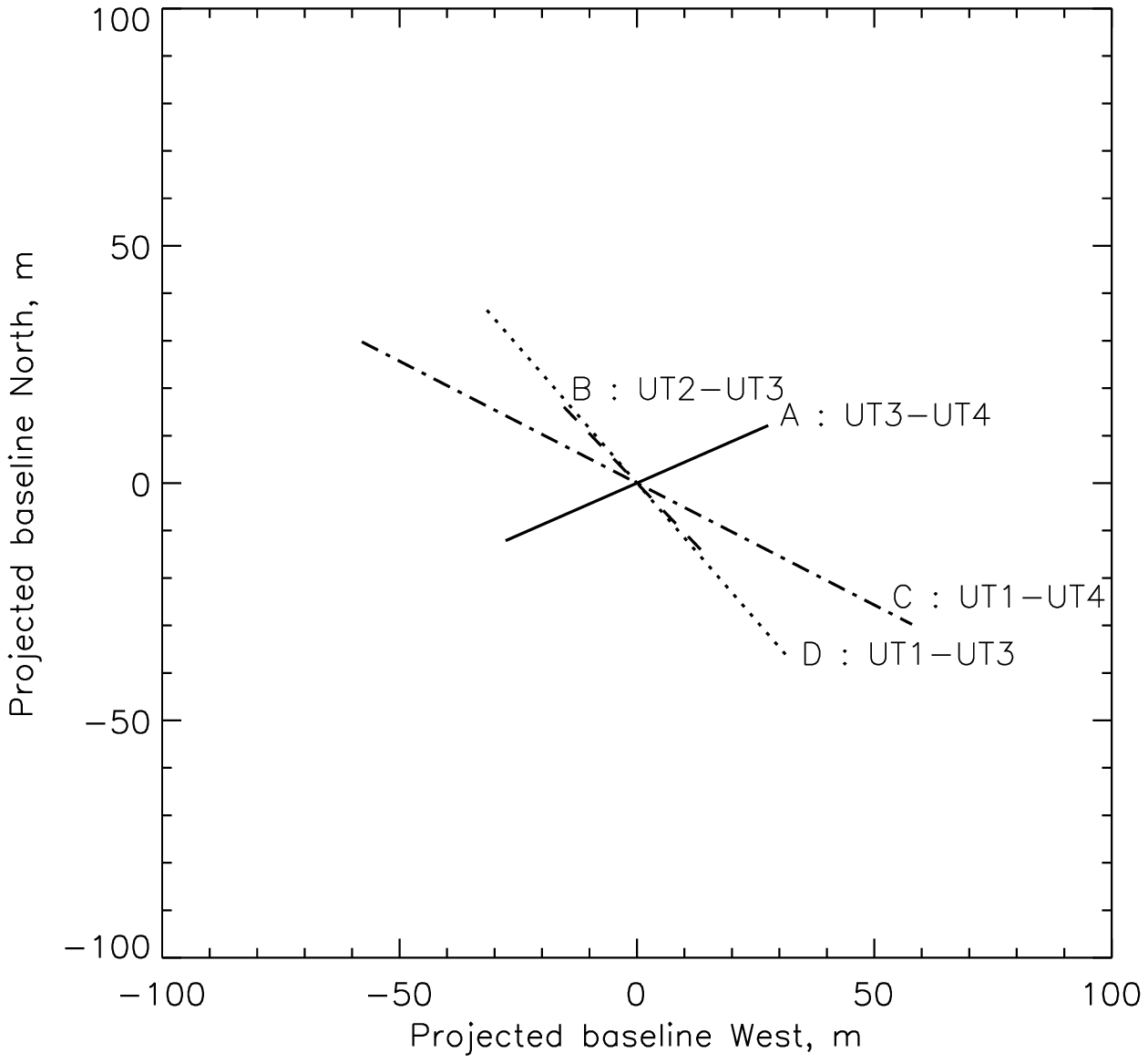}
\end{minipage}
\begin{minipage}{8cm}
\includegraphics[width=8cm]{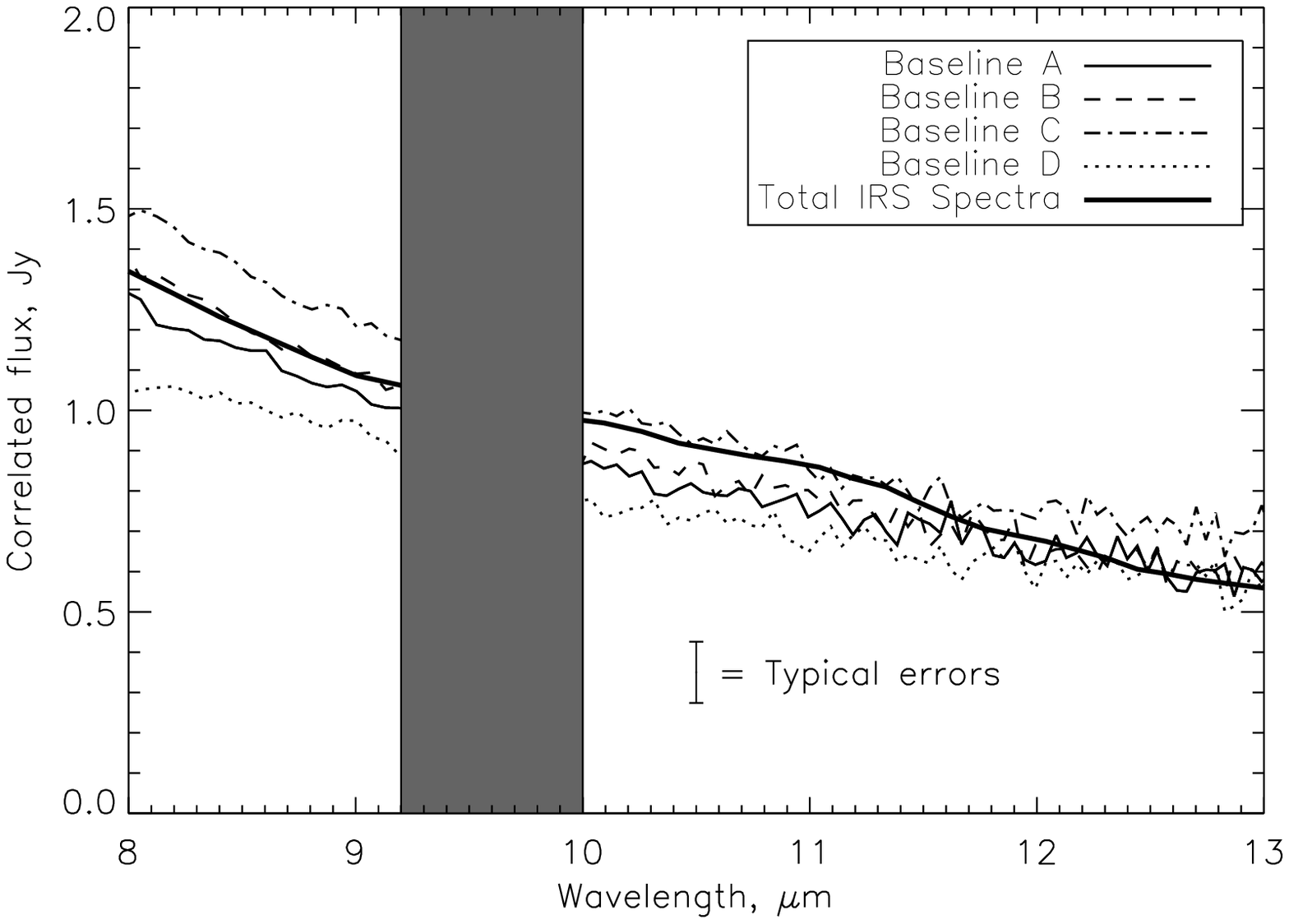}
\end{minipage}\\
\begin{minipage}{8cm}
\includegraphics[width=8cm]{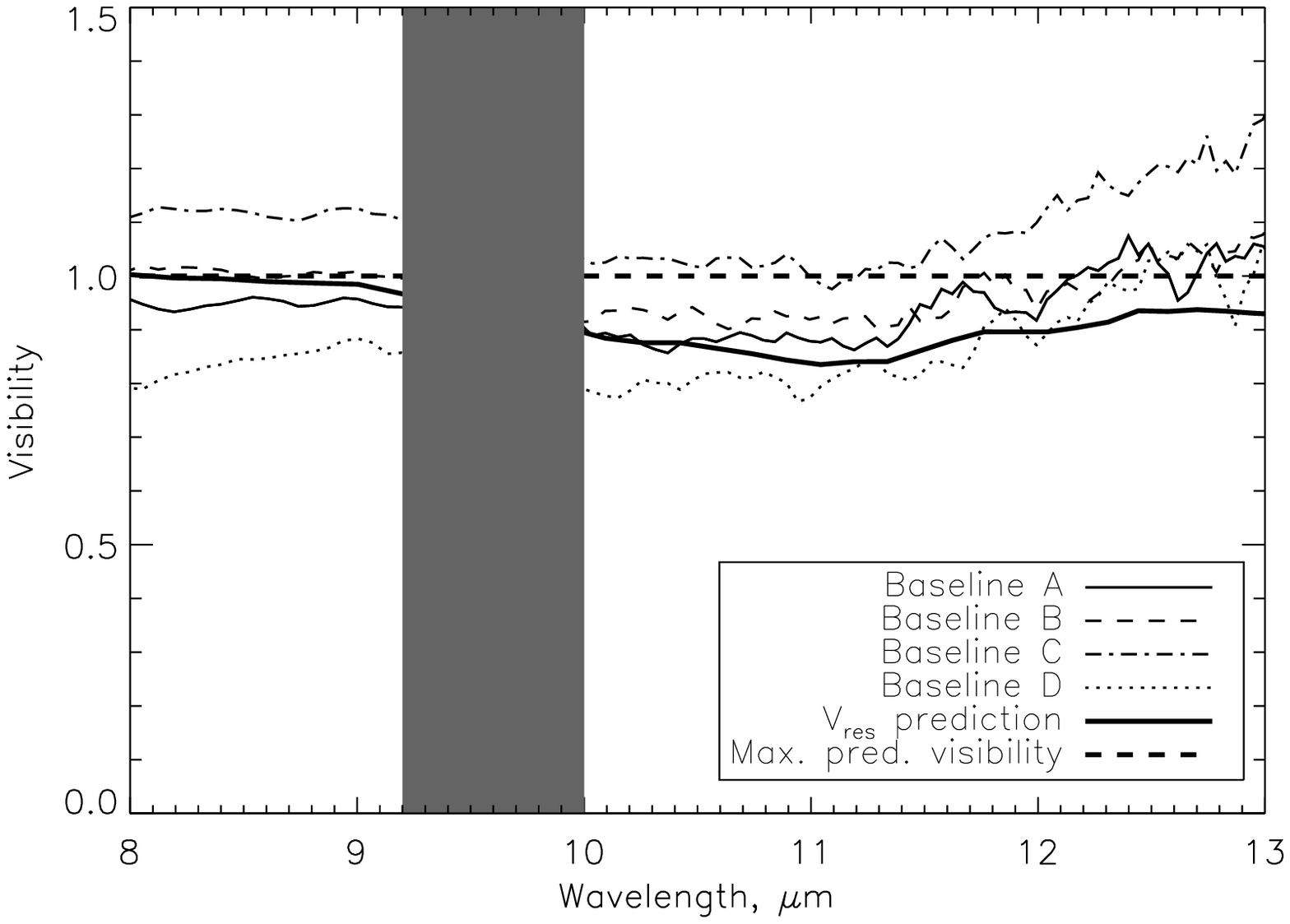}
\end{minipage}
\begin{minipage}{8cm}
\includegraphics[width=8cm]{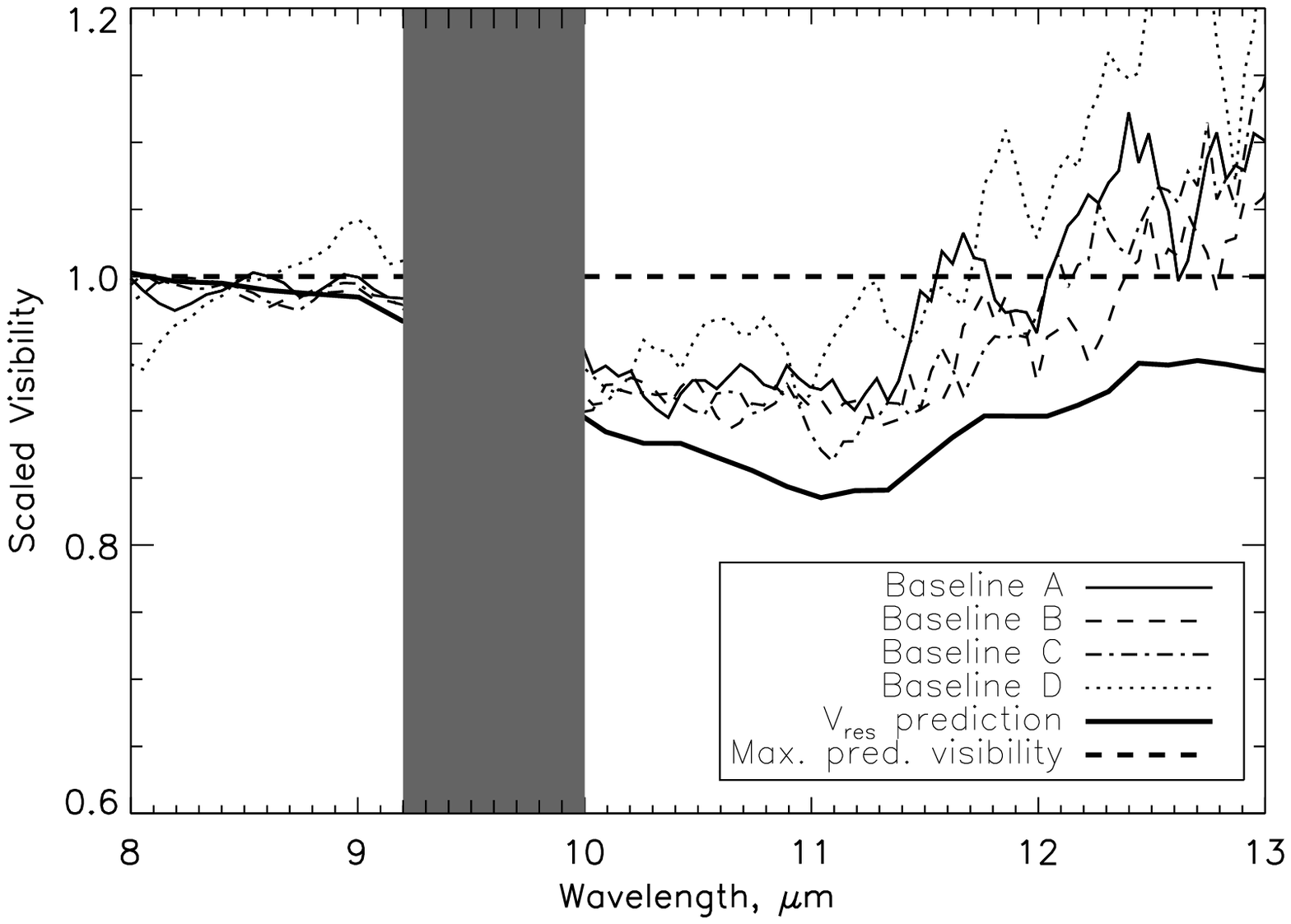}
\end{minipage}\\
\caption{\label{fig:69830obsres} The results of MIDI observations of
  HD69830. Top left: The baselines observed in this program. Top
  right: The calibrated correlated fluxes found in the
  MIDI observations of HD69830. The thick line represents the total
  IRS spectra, which represents the maximum flux we should see. Note
  that although the absolute values are varying by $\sim$ 10\%, all
  runs show a fairly flat spectrum, with no evidence of the silicate
  features seen in the IRS spectra. Bottom left: The visibility of
  HD69830. The measured visibilities on all four baselines are
  calculated using the IRS spectra and the measured correlated fluxes.
  The prediction for what the visibility curve should look like if the
  disc flux is totally resolved ($V_{\rm{res}}$, equation \ref{eq:vis_res})
  is also shown by a thick solid line. Bottom right: The visibility of
  HD69830 scaled to an average of 1 between 8-9 $\mu$m. Notice the
  shape across all baselines is similar to the $V_{\rm{res}}$
  prediction.}
\end{figure*}

\section{MIDI results}
\label{s:results}

We now present the results for each of the science targets in turn.
To help interpret our results, we consider how the
visibility function of a source is expected to behave if the source is
made up of multiple components. HD69830 and $\eta$ Corvi have
significant stellar emission in  
the N band (see Table \ref{tab:sources}).  Thus there will be two 
components to the visibility function, the component from the stellar 
emission (where the visibility $V_\star =1$ as the predicted angular
sizes of the stars are 0.63mas for HD69830 and 0.73mas for $\eta$
Corvi based on typical radii for their spectral types and their
distances determined by parallax) and the component from the disc
(where the $V_{\rm{disc}}$ will depend on the resolution of the disc
flux). The total visibility of a star + disc source will then be given
by 
\begin{equation}\label{eq:calcvis}
\qquad V_{\rm{tot}} = \frac{F_\star}{F_{\rm{tot}}} V_\star +
\frac{F_{\rm{disc}}}{F_{\rm{tot}}} V_{\rm{disc}}, 
\end{equation}
(where $F_\star$ is the flux of the star, etc.).
As only the disc components of our targets have the potential to be 
resolved on the baselines used in this study the 
equation above predicts a visibility function comprising the sum of two 
terms. The first of these will vary solely on the ratio of stellar to 
total flux whereas the second will depend both on the ratio of disc to 
total flux \emph{and} on whether the disc is resolved by the 
interferometric baseline. More precisely, the 
visibility function will tend to the value of 
\begin{equation}\label{eq:vis_res}
V_{\rm{tot}} \qquad  {\longrightarrow \atop V_{\rm{disc \rightarrow 0}}} \qquad
  V_{\rm{res}} = \frac{F_\star}{F_{\rm{tot}}} 
\end{equation}
when the 
disc is fully resolved.  As can be seen in Figure
\ref{fig:seds} the value of
$V_{\rm{res}} = \frac{F_\star}{F_{\rm{tot}}}$ is expected to change across the
spectral range of MIDI, particularly in respect of the silicate
features seen in the excess emission around both stars. As a result, 
any variation in observed visibility as a function of wavelength will 
need to be corrected for this behaviour before it can be interpreted 
in terms of any resolved source structure.

\subsection{HD69830}
HD69830 was observed on all four baseline combinations used for this
study.  A sketch of the geometry of the baselines and
relative lengths is shown in Figure \ref{fig:69830obsres} (top left).  

The calibrated correlated flux for all four baseline observations of
this source are shown in Figure \ref{fig:69830obsres} (top right).
The error bar indicates an average error measured across all four
baselines.  These errors are determined from two sources: splitting
the fringe tracking files into 5 equal subsets and performing the
reduction on each subset in turn to determine the variation in fringe
signal across the time of the observation (statistical error); and the
variation between correlated flux measurements as calibrated by
different standard star observations where available. The two sources
were added in quadrature.  The error is dominated by calibration error
which is typically $\sim$10\% of the correlated flux (see also 2.4). 

The correlated fluxes show variation across the different baselines,
as can be seen in the levels of flux measured at 8 $\mu$m. When
correlated fluxes are different on different baselines there is the
possibility that these differences represent real differences in the
resolution of the source.  However, at 8 $\mu$m the disc emission is
expected to be very low (4 mJy at 8$\mu$m), and so $\sim$ 100\% of the
emission at $8\mu$m should come from the stellar photosphere.  As the
star is expected to be point-like and completely unresolved, the
correlated flux should equal the total flux (marked by a thick solid line) at
this wavelength. The differences in these absolute values are
at the expected level of variation in absolute correlated flux from
changing conditions between science and standard star observations
(see section 2.4).  

What is clear from Figure \ref{fig:69830obsres} (top right) is that
for all baselines the correlated flux is flatter than the total flux
measured in the IRS spectrum; there is no evidence of the silicate
feature seen between 10--11.5 $\mu$m in Figure \ref{fig:seds}. As
shown by equation \ref{eq:vis_res}, if the disc flux is totally resolved
($V_{\rm{res}}$) the
correlated flux should consist solely of the stellar photospheric
emission - with no evidence of the silicate features seen in the
excess emission spectrum.  The indication that the disc flux has been
at least partially resolved on all baselines is made clearer by
examination of the visibility function shown in Figure
\ref{fig:69830obsres} (bottom left). 

The absolute values of visibility across all baselines are an average
of 1.04$\pm$0.16. That this variation is a reflection of absolute
uncertainty and not evidence of different levels of resolution is
confirmed by the visibility at 8 $\mu$m, which as discussed above
should be 1 as the flux here is completely dominated by the stellar
photosphere.  Uncertainty at this level is to be expected,
given typical uncertainties in calibrated visibilities of 0.10
(see Figure \ref{fig:calcalvis}) and the relative faintness of the
target (total flux of $\sim$1 Jy across the MIDI spectral range, see
Figure \ref{fig:seds}).  Beyond this uncertainty in absolute values,
however, the visibilities measured on all four baselines can be seen
to dip exactly in the regions we would expect a dip if the disc flux
was resolved. This dip is made clearer if we scale the visibility
function to 1 averaged over the short wavelength region (8--9$\mu$m),
as would be expected even in the resolved disc case ($V_{\rm{res}}$)
because of the lack of excess in this region.

\begin{table}
\begin{tabular}{*{4}{|c}|}\hline Baseline &
  \multicolumn{3}{|c|}{Visibility} \\ name & 8-9$\mu$m & 10-11.5$\mu$m
  & Ratio   \\ \hline
A & 0.944 $\pm$ 0.003 & 0.879 $\pm$ 0.005 & 0.931 $\pm$ 0.037 \\
B & 1.003 $\pm$ 0.003 & 0.928 $\pm$ 0.004 & 0.926 $\pm$ 0.027 \\
C & 1.093 $\pm$ 0.003 & 1.018 $\pm$ 0.004 & 0.931 $\pm$ 0.028 \\
D & 0.839 $\pm$ 0.006 & 0.803 $\pm$ 0.005 & 0.958 $\pm$ 0.039 \\ \hline
$V_{\rm{res}}$ & 0.990 & 0.860 & 0.869  \\ \hline
\end{tabular}
\caption{\label{tab:69830visdip} Measuring the significance of the dip
  in visibility of HD69830 at 10-11.5 $\mu$m. The values of visibility
  in each wavelength range are the weighted means taken over all spectral
  channels in that range, and uncertainties quoted are the standard
  errors on the weighted mean from statistical uncertainty only
  (uncertainty in the absolute visibility is higher due to calibration
  uncertainties). Final
  error on ratio includes error from the calibrator ratios discussed
  in section 2.4 and a 1.3\% error on the spectral slope of IRS
  observations \citep{beichman06}, as well as the statistical errors
  from each wavelength range.  The $V_{\rm{res}}$
  predicted visibility arises from assuming that $V_{\rm{disc}}=0$
  (see equation \ref{eq:vis_res}). }
\end{table}

%
The scaled visibility functions measured on the four baselines are
shown in Figure \ref{fig:69830obsres} bottom right. The dip in
visibility between 10-11.5 $\mu$m is now very obvious.  To confirm the
significance of this dip, the unscaled average visibility within the
14 spectral channels between 8-9 $\mu$m was compared to that in the 29
spectral channels between 10-11.5$\mu$m (i.e., around the lowest
predicted visibility when $V_{\rm{disc}} = 0$). Table
\ref{tab:69830visdip} presents these weighted means (the weights were
determined from the statistical variation within the 5
sub-integrations available for each spectral channel) together with
their standard errors.  Calibration uncertainty is not included in
these values. The visibility ratios presented in the table are the
ratio of these weighted means, where we have included a contribution
from calibration errors in quadrature with the statistical errors in
the 2nd and 3rd columns when evaluating the errors in the final
column.  These calibration errors arise from two sources.  First, as
we have used the IRS spectrum to determine the visibilities, we
included an error of 1.3\% typical of errors on the IRS spectral slope
as measured for sources with no observed excess emission
\citep{beichman06}. Second, we included the error on the ratio from
calibration as measured in the standard--standard calibrated
visibilities discussed in section 2.4.  We used the standard deviation
of the visibility ratio (3.7\%, section 2.4), but for runs B and C
where 2 standard star observations were available for calibration,
this was reduced by a factor of $\sqrt{2}$.
%

%
For all four baselines used to observe HD69830 the visibility ratio
suggests a partial resolution of the disc - the ratio is higher than
the 0.869 expected for $V_{\rm{res}}$ but definitely lower than
unity. On baselines B and C this difference from unity is significant
at a level of $\sim$ 3$\sigma$. When averaged over all 4 baselines the
weighted mean of the visibility ratio is 0.934 $\pm$
0.015.\footnote{\label{foot} Note that 
  these calculations assume that the visibility on all baselines is
  the same.  Although this is not generally expected to be the case it
  is true if the disc is completely resolved or unresolved and so
  these calculations are used to determine the significance with which
  we can say that the disc is completely resolved or unresolved.}
Taken as a whole, 
these data suggest that the observed visibility ratio is different
from 1 at a level of 4$\sigma$.  We note
in passing that the actual scatter between the four measurements of
the visibility ratio is very small, suggesting that we have been
conservative in associating an error of $0.015$ to our final
``best-estimate''.

We do note, however, that the lower right panel of
Figure \ref{fig:69830obsres} also shows the visibility function of
HD69830 rising beyond $\sim$12$\mu$m to above 1 in the scaled case,
meaning the ratios between the weighted mean visibilities between
11.5--13$\mu$m and between 8--9$\mu$m are on average greater than 1
(baseline A 1.045, B 0.991, C 1.043 and D 1.098).  For baseline D this
ratio is larger than would be expected from any calibration error, as the
error on the ratios over the same wavelength ranges from the
calibrator-calibrator visibilities (section 2.4) is a maximum of
0.053.  The $V_{\rm{res}}$ prediction shows that the visibility
function in the case that the disc is completely resolved would be
expected to rise, but visibilities should not ever be greater than 1
(see equation \ref{eq:vis_res}). This is unlikely to be due to the
water ice discussed in section 2.3, as if some of the excess emission
measured in the IRS spectrum did fall outside of the MIDI beam we
would be over-estimating $I_{\rm{tot}}$ in our calibration, and this
would lead to a falsely low visibility.  We consider that the lower
signal-to-noise in the MIDI fringe measurements on this target at
longer wavelengths (signal-to-noise is an average of $\sim$70 between
8--9$\mu$m, 37 between 10--11.5$\mu$m and 15 beyond 12$\mu$m) is the
cause of these high visibilities which causes a residual
slope. Whether this affects the 10--11.5$\mu$m region is not clear but
for this reason we do not consider that the results provide strong
evidence for a partially resolved rather than completely resolved
disc.  
%
%
%
%
%
The limits we can place on the morphology of
the HD69830 debris disc with the results presented in this section are
discussed in section 5.1.

\subsection{$\eta$ Corvi}

\begin{figure*}
\begin{minipage}{8cm}
\includegraphics[width=8cm]{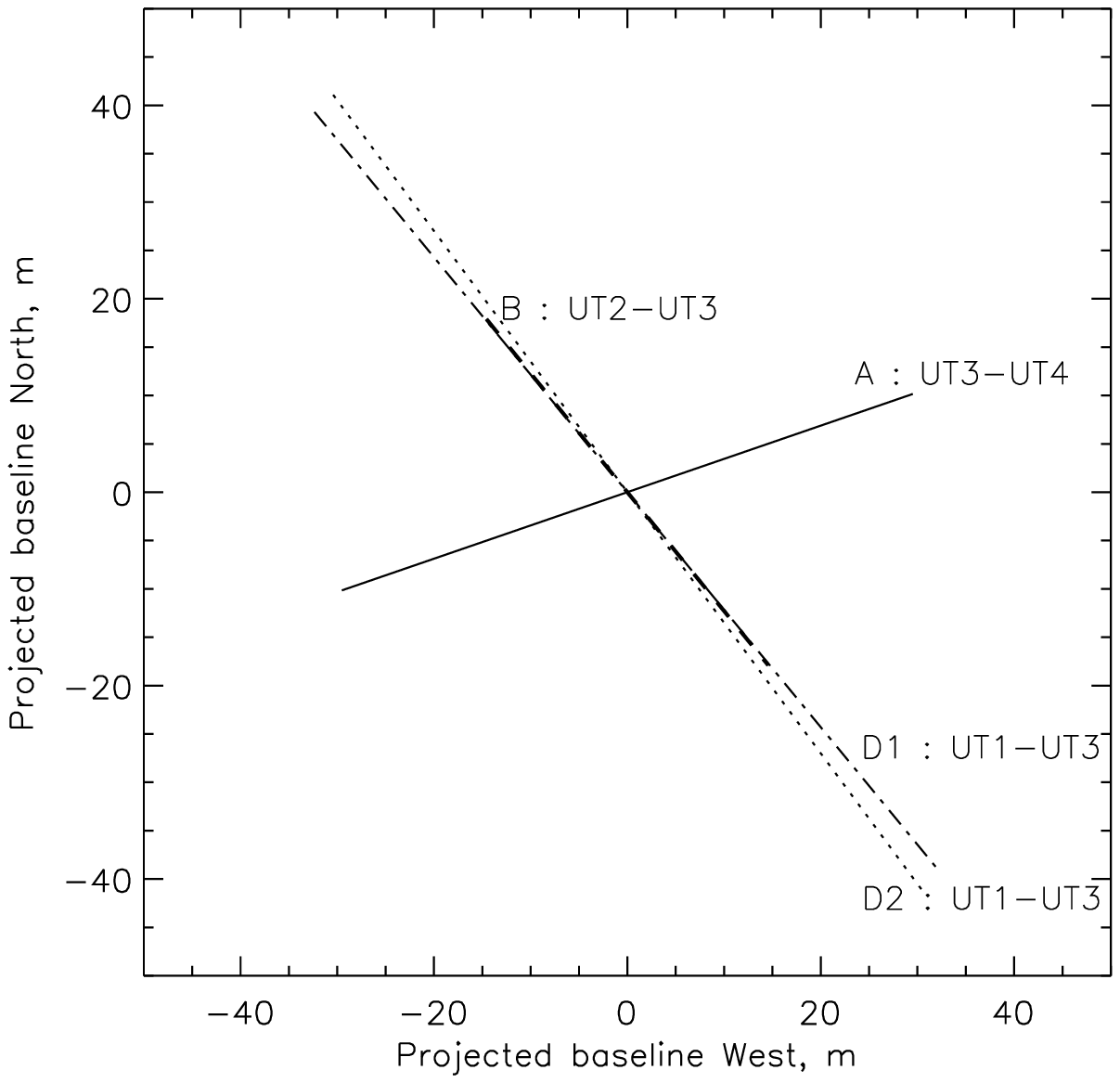}
\end{minipage}
\begin{minipage}{8cm}
\includegraphics[width=8cm]{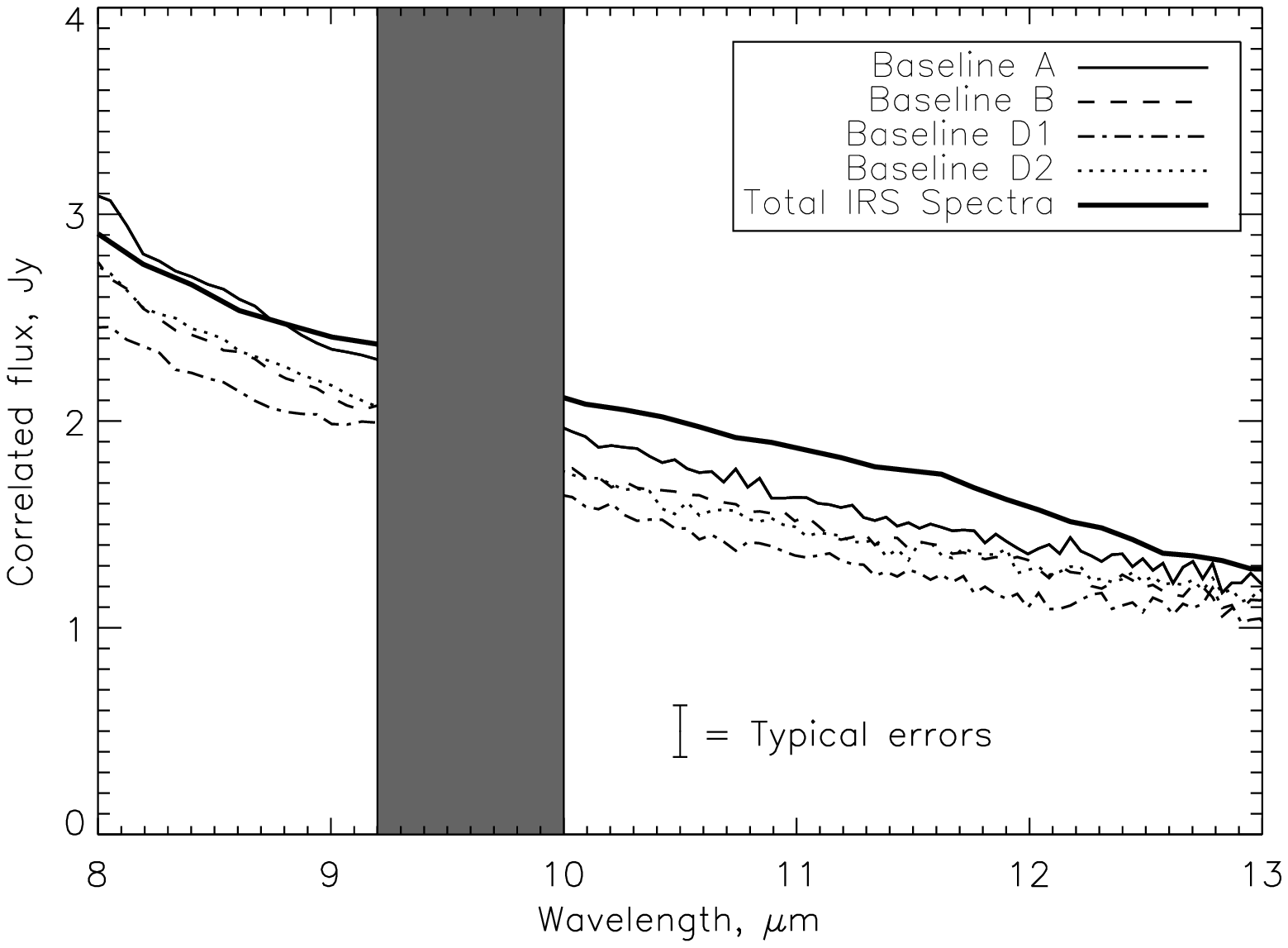}
\end{minipage}\\
\begin{minipage}{8cm}
\includegraphics[width=8cm]{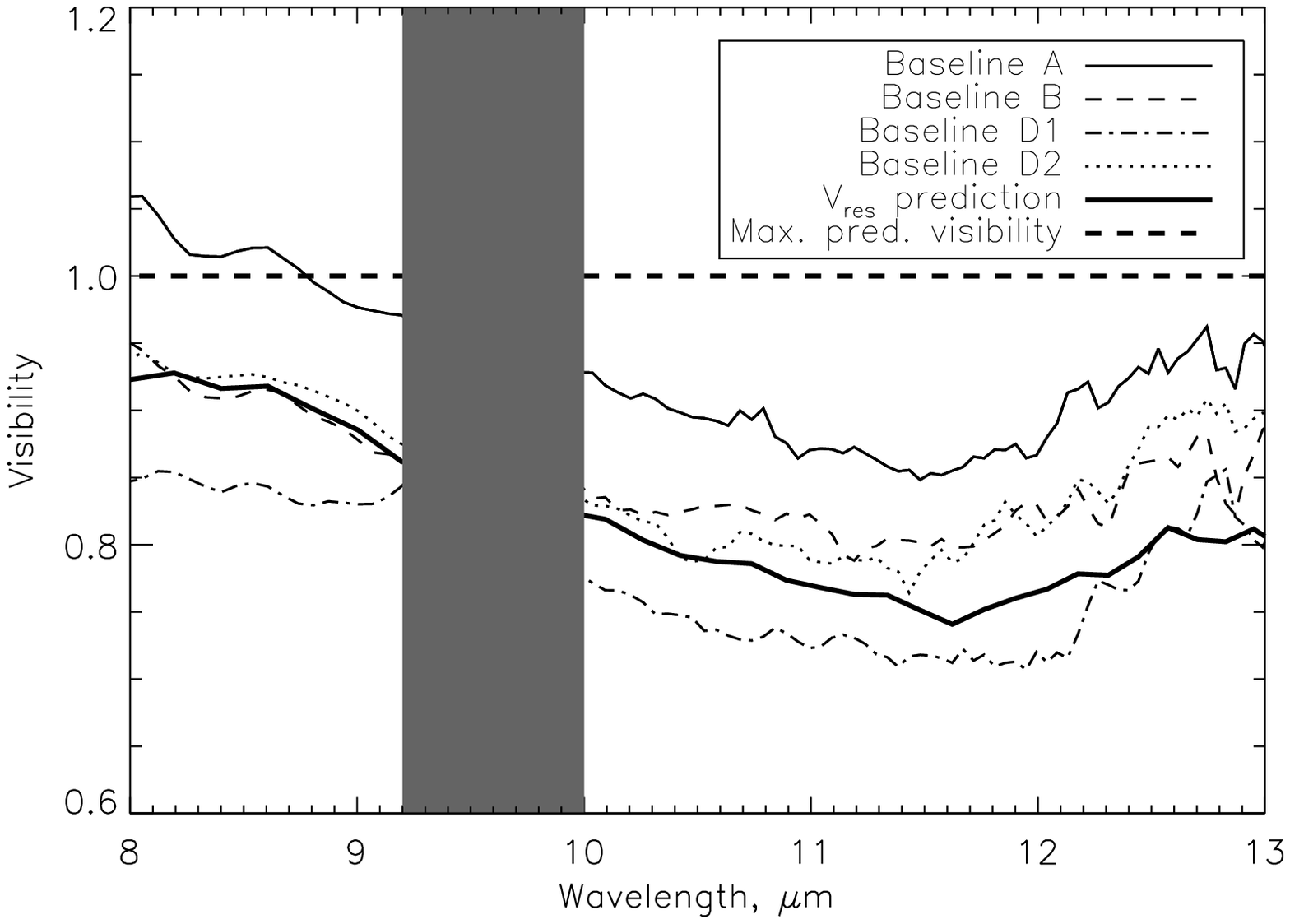}
\end{minipage}
\begin{minipage}{8cm}
\includegraphics[width=8cm]{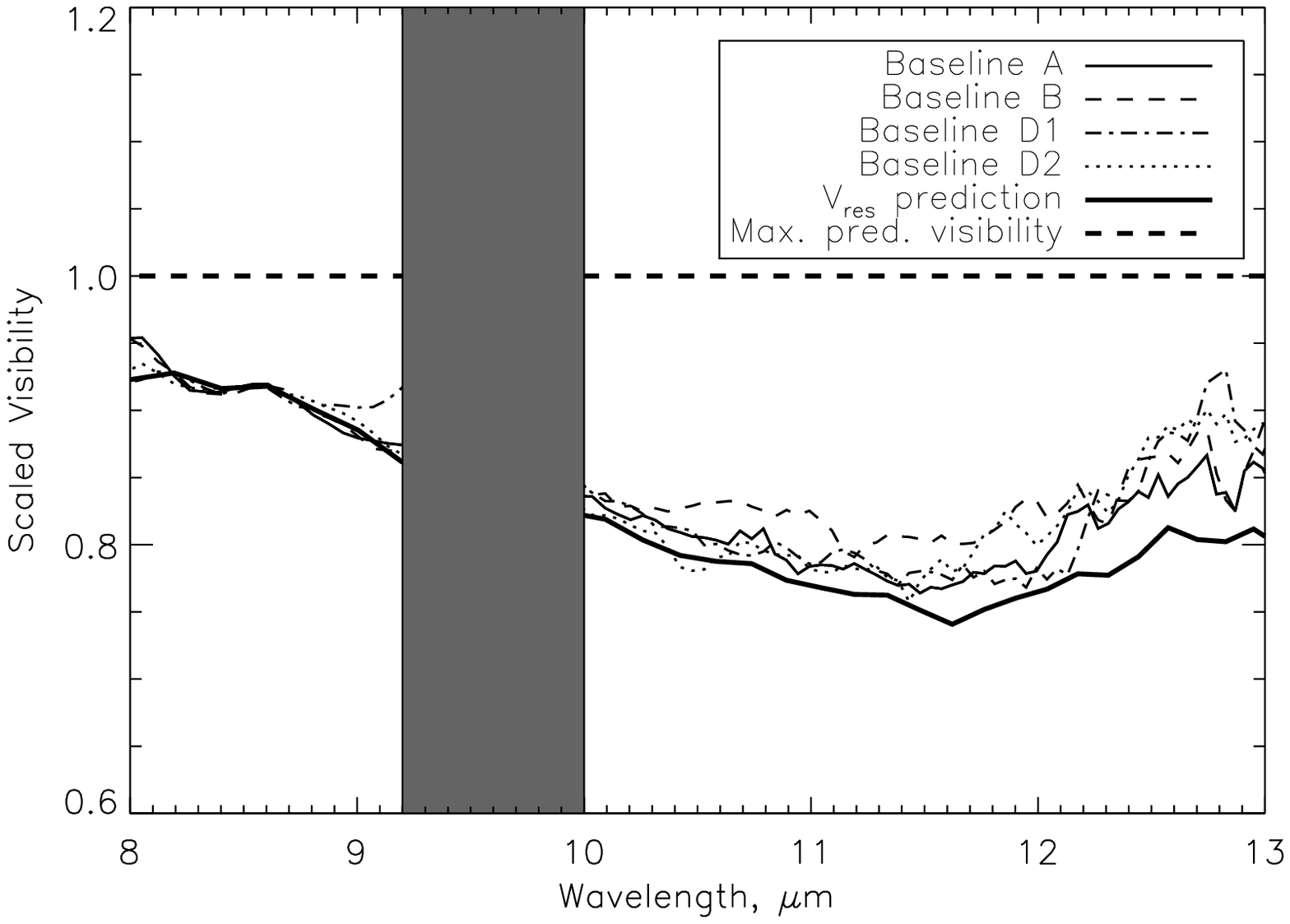}
\end{minipage}\\
\caption{\label{fig:etaobsres} The results of MIDI observations of
  $\eta$ Corvi. Top left: The baselines observed in this program. Top
  right: The calibrated correlated fluxes found in the
  MIDI observations of $\eta$ Corvi. The thick line represents the total
  IRS spectra, which represents the maximum flux we should see. Note
  that although the absolute values are varying by $\sim$ 10\%, all
  runs show a fairly flat spectrum, with no evidence of the silicate
  features seen in the IRS spectra. Bottom left: The visibility of
  $\eta$ Corvi. The measured visibilities on all four baselines are
  calculated using the IRS spectra and the measured correlated fluxes.
  The prediction for what the visibility curve should look like if the
  disc flux is totally resolved ($V_{\rm{res}}$, equation \ref{eq:vis_res})
  is also shown by a thick solid line. Bottom right: The visibility of
  $\eta$ Corvi scaled to the average visibility from the
  $V_{\rm{res}}$ prediction between 8-9 $\mu$m. This scaling shows the
  four observations all show strong evidence of being similar to the
  $V_{\rm{res}}$ prediction.}
\end{figure*}

%
%
$\eta$ Corvi was observed on baselines A and B and on baseline D twice
(see Figure \ref{fig:etaobsres}, top left panel).  The second
observation on baseline D was carried out with an increased chop
frequency and integration time which led to greatly improved intensity
calibration data.

The calibrated correlated fluxes for all baselines are shown in Figure
\ref{fig:etaobsres}, top right, where we have included calibration
errors and the variation across sub-exposures added in quadrature.
The error bar shown is indicative of the average error observed over
all baselines. Overplotted on these figures is the total IRS
photometry presented in \citet{chen06}, as also shown in Figure
\ref{fig:seds}.
%
%

The correlated fluxes are very similar in shape across all four
observations. The absolute values vary by the typical $\sim$10\% we
expect for absolute variation in correlated flux (see section 2.4),
but all show a spectrum consistent with photospheric emission in the
Rayleigh-Jeans regime.  In common with the result of HD69830, there is
no evidence of the spectral feature seen in the IRS total spectrum in
any of the correlated flux measurements.  This indicates that for this
source the disc emission also appears to be partially resolved. Again,
the result is made clearer by examination of the visibility function.

The calibrated visibilities for all observations of $\eta$ Corvi are
shown in Figure \ref{fig:etaobsres} (bottom left),
along with the visibility predicted for a completely resolved disc
component marked in a thick solid line ($V_{\rm{res}}$, see equation
\ref{eq:vis_res}). The dip in visibility at $\sim$11.5 $\mu$m is
seen very clearly on all observed baselines.  Again, as expected, 
the variations between different baselines in terms of absolute values are
an average of $\sim$0.1, with a maximum difference of $\sim$0.2.
%
%
%
%
The visibilities scaled to the $V_{\rm{res}}$ prediction between
8--9$\mu$m are shown in Figure \ref{fig:etaobsres} (bottom
right). Although we do not know what the visibility should be at
8--9$\mu$m for $\eta$ Corvi (as we do for HD69830) as there is
significant disc flux in this range, this scaling allows the shape of
the visibility function to be seen more clearly and allows easy comparison with
the $V_{\rm{res}}$ model.  

In Table \ref{tab:etadip} the weighted mean visibilities in the short
wavelength region (8--9 $\mu$m) and the region in which the silicate
excess feature has the strongest contribution to the total flux
(10--11.5 $\mu$m) are listed.  Standard errors and errors on the ratio
including calibration uncertainty are calculated as described for
HD69830 in section 3.1.  As is clear from the table the
difference seen in the measured  visibilities at
shorter wavelengths and midway through the MIDI range (10-11.5$\mu$m)
is significant. Not only this, but the size of the change is similar
in all cases to that predicted for the case of completely resolved
excess emission, as can be seen in the scaled visibility function plots
(Figure \ref{fig:etaobsres} bottom right). Results on individual
baselines are incompatible with an unresolved disc (ratio = 1) at a
level of at least 4 $\sigma$ (see Table \ref{tab:etadip}).  Taking all
the results together (see footnote \ref{foot}) the weighted mean dip in
visibility is 0.880 $\pm$ 
0.013, incompatible with an unresolved disc at the $9 \sigma$ level.
The individual measured ratios and average result are close to
the prediction for a completely resolved disc ($V_{\rm{res}}$). Note
that we do not see the same evidence for a significant residual slope
in the visibility function as was the case for HD69830.  The
observations of $\eta$ Corvi benefit from a higher signal-to-noise
even at longer wavelengths (average signal-to-noise in the correlated
flux measurements is 128 at 8--9$\mu$m, 75 at 10--11.5$\mu$m and
31 at wavelengths beyond 12$\mu$m). 
Though the levels of absolute calibrated flux and thus visibility are
not constrained to better than 10\%, (see section 2.4) the dip and its
appearance at a significant level across all baselines is consistent
with the disc emission being completely resolved on all observed
baselines.  The constraints we can place on the morphology of this
source based on these results are discussed in section 5.2. 

\begin{table}
\begin{tabular}{*{4}{|c}|}\hline Baseline &
  \multicolumn{3}{|c|}{Visibility} \\ name & 8-9$\mu$m & 10-11.5$\mu$m
  & Ratio  \\ \hline
A & 1.016 $\pm$ 0.005 & 0.892 $\pm$ 0.004 & 0.877 $\pm$ 0.026 \\ 
B & 0.910 $\pm$ 0.003 & 0.818 $\pm$ 0.003 & 0.899 $\pm$ 0.027 \\
D1 & 0.841 $\pm$ 0.003 & 0.739 $\pm$ 0.003 & 0.879 $\pm$ 0.026 \\
D2 & 0.922 $\pm$ 0.003 & 0.801 $\pm$ 0.004 & 0.868 $\pm$ 0.026 \\ \hline
$V_{\rm{res}}$ & 0.915 & 0.784 & 0.857  \\ \hline
\end{tabular}
\caption{\label{tab:etadip} Measuring the significance of the dip
  in visibility of $\eta$ Corvi at 10-11.5 $\mu$m.  A description of
  the calculation of the visiblities and errors is given in the
  caption to Table \ref{tab:69830visdip}. }
\end{table}


\section{VISIR imaging of HD69830}

\begin{figure*}
\begin{minipage}{6cm}
\center{Standard} \\
\includegraphics[width=6cm]{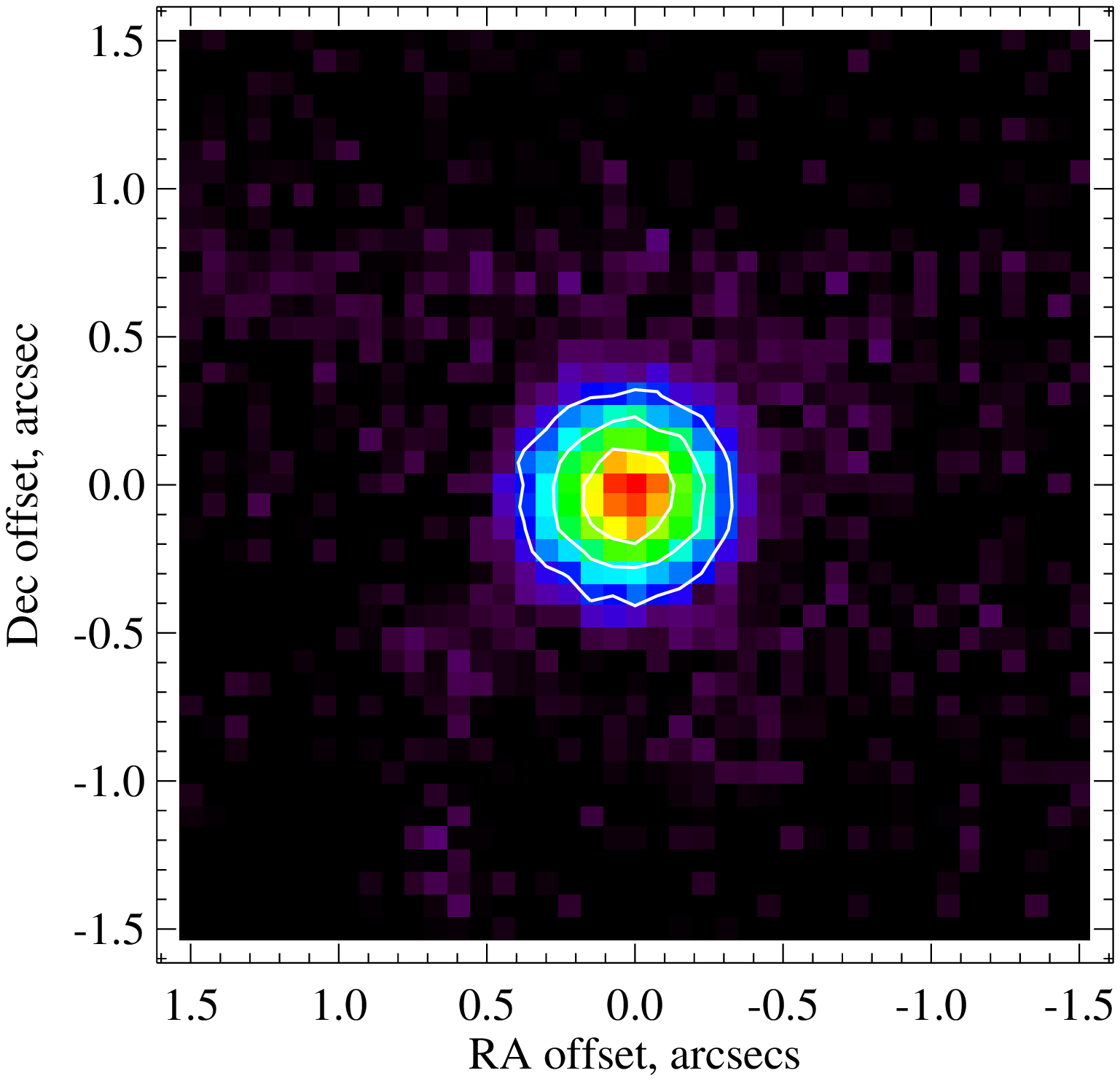}
\end{minipage}
\begin{minipage}{6cm}
\center{HD69830} \\
\includegraphics[width=6cm]{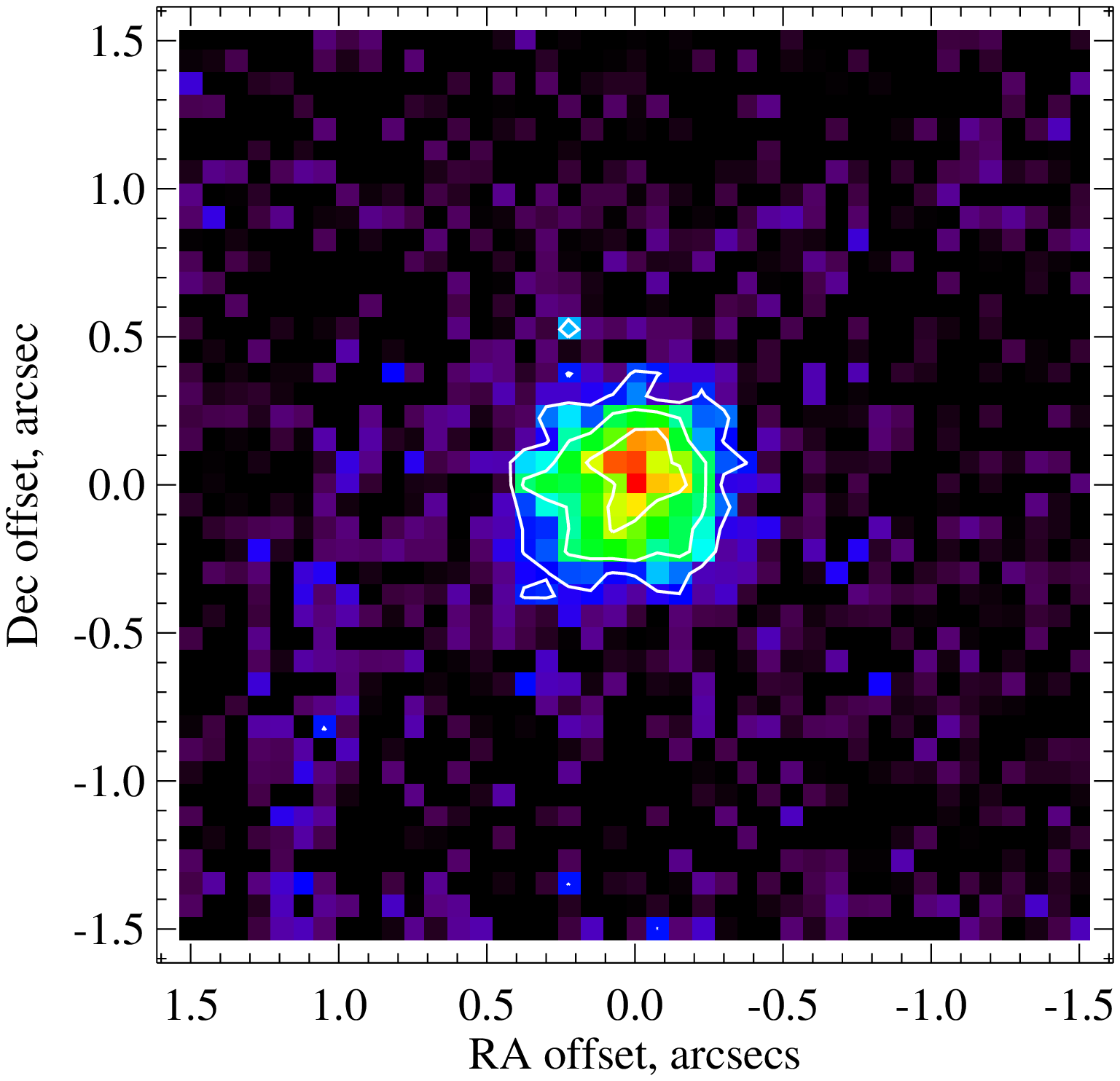}
\end{minipage}
\begin{minipage}{6cm}
\center{Residuals} \\
\includegraphics[width=6cm]{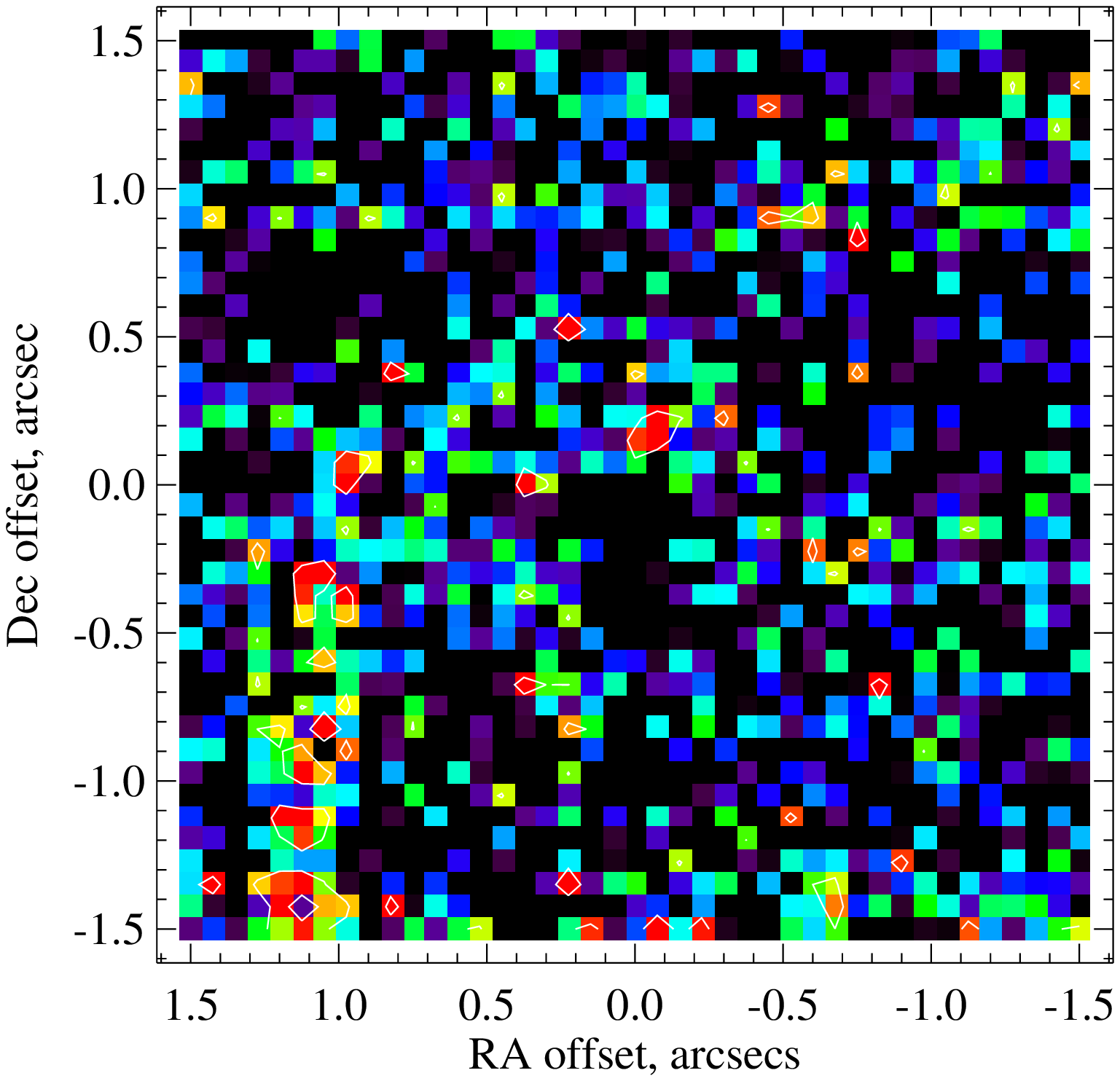}
\end{minipage}
\caption{\label{fig:visir69830}The VISIR images of HD69830 and
  standard star observed in filter Q2.  Left: The
final image of standard star (HD61935) after co-addition of 4
sub-integrations. This image acts as a PSF reference. Middle: The
final image of HD69830.  Contours on this and the standard star are at
levels of 25, 50 and 75\% of the peaks.  No evidence of emission
extended beyond the PSF reference is seen in the image of
HD69830. The colour scale on these images goes from 0 to the peak. 
Right: The image of HD69830 after subtraction of the PSF
reference (standard image) scaled to the peak. The colour scale has a
minimum of 0 and maximum of 3$\sigma$ per pixel. Contours mark any
region of $>$2sigma significance. The stripe to the South East of the
image is a detector artifact.  No significant emission is seen after
subtraction of the PSF which would indicate resolved emission
(testing over a range of regions of different shapes which could
detect a range of different disc geometries gives a maximum
detection of $<$2$\sigma$ in this image). }
\end{figure*}

Following the 8m limits on the $\eta$ Corvi emission presented 
in \citet{smithhot}, we have new data obtained on VISIR at the 
VLT under proposal 079.C-0259 on 6$^{\rm{th}}$ April 2007 on the
emission around HD69830. The data 
were all taken in filter Q2 ($\lambda_c = 18.72\mu$m, 
$\Delta\lambda = 0.88\mu$m) with pixel scale 0\farcs075. The
observations  
were performed using a chop and nod throw of 8\arcsec.  Chopping was 
performed in a North-South direction and nodding was performed in 
the perpendicular direction.  The observations consisted of two 
integrations of 2100s each on HD69830, with standard star observations
before and after each science target integration (125s per
integration).  Frequent standard star 
observation was used to determine how the PSF was varying through 
the observations.  The standard star observations were also used for 
photometric calibration.  The standard star, HD61935, was chosen from 
a list of mid-infrared spectro-photometric standards \citep{cohen} and
was also used in the MIDI observations (Table \ref{tab:sources}).
The observations are summarised in Table \ref{tab:visirobs}. 

The data were reduced using custom routines described in detail in 
\citet{smithhot}.  In summary, data reduction involved determination 
of a gain map using the mean values of each frame to determine pixel 
responsivity (masking off pixels on which emission from the source 
could fall, equivalent to a sky flat).  In addition a dc-offset was
determined by calculating the mean pixel values in columns and rows
(excluding pixels on which source emission was detected) and this was
subtracted from the final image to ensure a flat background. Pixels
showing high or low gain, or those which showed great variation
throughout the observation, were masked off. The chop-and-nod pattern
adopted results in four images of the star falling on the detector.
These images were co-added after determining the centre of each image
by 2-dimensional Gaussian fitting and then the sub-integrations were
added together to give a final image of HD69830 and the standard star
which would act as a PSF reference.  

\begin{table}
\centering
\begin{tabular}{*{3}{|c}|} \hline Target & Target type & Integration
  time (s) \\ \hline 
HD61935 & Cal & 125 \\ HD69830 & Sci & 2100 \\ HD61935 & Cal & 125 \\
HD61935 & Cal & 125 \\ HD69830 & Sci & 2100 \\ HD61935 & Cal & 125
\\ \hline 
\end{tabular}
\caption{\label{tab:visirobs} The observations of HD69830 and standard
  star with VISIR on the VLT.  Observations were performed under
  proposal 079.C-0259.  All observations were performed using filter
  Q2 and the small field detector  (pixel scale 0\farcs075).  }
\end{table}

The photometry was performed using circular apertures of 1\arcsec
radius centred on the peak of the emission.  Statistical background
noise was  
calculated in annuli centred on the peak of emission with inner
radius 2\arcsec, outer radius 4\arcsec.  Calibration uncertainty was
estimated from the variation in calibration levels from the four
sub-integrations on the standard star, and was found to be 9\%.
HD69830 was detected in the final co-added image with a
signal-to-noise of 32, with the total photometry including 
calibration uncertainty of 377 $\pm$ 46 mJy. The IRS spectrum of 
this source has a flux of 365 mJy at 
18.72$\mu$m (253 mJy expected from the photosphere). 
No colour-correction was applied for this narrow-band filter.  In
addition, no correction for airmass was employed, as the standard star
was observed at a similar airmass to HD69830 and no evidence was found
that the calibration factors from the sub-integrations on the standard
star were correlated with airmass. 

The final images of the standard star HD61935, used as a PSF
reference, and of HD69830 are shown in Figure \ref{fig:visir69830}.
Overplotted on the images are contours at 25, 50 and 75\% of the image
peak (brightness peaks were 7980 mJy/arcsec$^2$ for the standard star
observation, 1033 mJy/arcsec$^2$ for HD69830).  Comparison of the
contour plots indicates that there is no sign of extended emission
around HD69830 that could be indicative of a resolved disc.  This is
confirmed by examination of the residuals image (Figure
\ref{fig:visir69830} right).  This is the HD69830 image after
subtraction of 
the PSF reference image (HD61935, Figure \ref{fig:visir69830} left)
scaled to the peak of emission.  The residual image shows no evidence
for emission beyond the PSF.  Further tests comparing the profiles of
the observations, Gaussian and Moffat profile fits to the images and
comparisons of FWHM measurements all confirmed this result. Different
test regions (annuli of different dimensions for face-on disc
geometries, rectangular boxes for edge-on discs, and intermediate
shapes for discs with intermediate inclinations to the line of sight)
were also examined for any evidence of significant emission in the
residuals image.  No such emission was found (most significant
positive flux of just less than 2$\sigma$ significance was found in a
region including a detector artifact - see Figure \ref{fig:visir69830}).

\begin{figure}
\includegraphics[width=8cm]{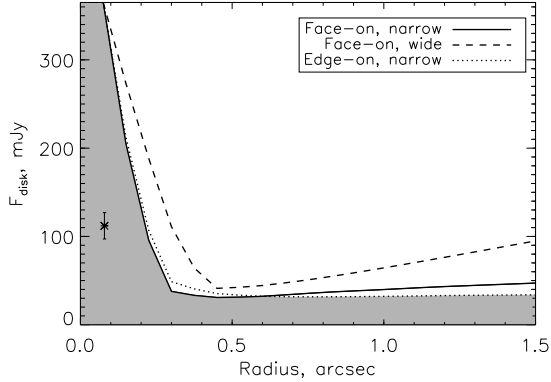}
\caption{\label{fig:visirlim} The limits we can place on the geometry
  of the HD69830 following no detection of extended emission in the
  VISIR Q band imaging. Discs in the space above the lines should have
  been detected in our imaging, and those lying in the shaded region
  would have remained undetected regardless of geometry.  Different
  lines represent different disc geometries as given in the legend. 
  Narrow discs have a disc width of $0.2r$, wide discs a width of
  $2.0r$, where $r$ is the radius of the mid-point of the disc.
  The expected disc flux and location from the
  SED is marked by an asterisk, with error bars representing the
  uncertainty in $R_{\lambda}$ from errors in the IRS spectrum. }
\end{figure}

We followed the procedure of \citet{smithhot} to place limits on the
size of the disc after no extension was detected.  This procedure is
described in detail in \citet{smithhot}, and involves convolving the
PSF model with different disc + star models to determine what
combinations of disc parameters (geometry and emission level) would
have resulted in a detection of extended emission.  The limits depend
strongly on the sensitivity of the observation and the variation of
the PSF, which was determined in this case from examination of the
four sub-integrations on the standard star. The PSF was found to have
a FWHM of 0\farcs52$\pm$0.04, a variation of 7\%.  
The individual integrations gave four
separate PSF models which were used in the limits testing, with
variations between the results being the limiting factor on disc
detection at small disc radii.  The results of this testing are
shown in Figure \ref{fig:visirlim}. For a face-on disc assuming a
narrow ring-like structure (width $0.2r$ where $r$ is the disc radius)
the extension testing places 
a limit of $<$0\farcs19 $\pm$ 0\farcs01 on the disc radius ($<$2.4 AU at a
distance of 12.6pc). This limit is determined by the size and
variation of the PSF (limiting line at small disc radii is near
vertical as it is not strongly dependent on the disc flux). A PSF
variation of 7\% is fairly typical for Q band observations taken under
good conditions based on previous observing experience \citep[see
  e.g.][]{smithhot,smitheta}.  Only
observations taken under optimal (very stable seeing) conditions would
allow us to probe closer to the star, and then only by perhaps one or
at best two tenths of an arcsecond ($\sim$0.1-0.25AU).  This limit
assumes disc emission is 112$\pm$15mJy (3$\sigma$ limit using errors
from IRS photometry), as measured in the IRS spectrum assuming a
Kurucz model profile for the photosphere and in
agreement with our photometry.  Limits on broader discs arise from the
sensitivity of the observations, and could be improved with increased
observation time.  With the current limits, we can rule out any disc
brighter than 33mJy at a radius of 0\farcs5 (6.3AU) with some
dependence on the assumed disc geometry (Figure \ref{fig:visirlim}). 
  

\section{Discussion}

We now consider the implications for the geometries of the HD69830 and
$\eta$ Corvi discs in the mid-infrared following the MIDI results and the
constraints from 8m imaging.

\subsection{HD69830}

The MIDI correlated fluxes measured for HD69830 and the 
  visibility functions determined from these are
  consistent with the detection of photospheric emission only.  The
  slopes are consistent with a Rayleigh-Jeans slope, and there is no
  evidence of the silicate feature observed in the IRS spectrum.
  We thus conclude that the excess emission does not strongly
  contribute to the correlated flux and therefore this has been at
  least partially resolved on the observed baselines.

Evidence for resolved emission around $\zeta$ Aql in the
near-infrared was interpreted as evidence for a low-mass companion object
\citep{absil08}.  We do not believe this is the explanation for the
resolved emission around HD69830, since the emission spectrum
is consistent with dust emission \citep{beichman05,lisse07},
and the observed visibility ratio is the same (within the errors) on
all observed baselines.
The same resolution on near perpendicular baselines such as A and B/D
(see Table \ref{tab:obs} and Figure \ref{fig:69830obsres} top left) would
not be expected if the emission arose from a point-like source.
Furthermore extensive radial velocity observations have revealed only
low-mass companions \citep{lovis} orbiting within 1AU, believed to
be Neptune-mass (although the masses could be larger if the system
is observed face-on).
Such planets cannot be the source of the observed change in
visibility with wavelength due to the low expected star:planet
mid-infrared flux contrast of $<10^{-3}$ \citep[see
e.g][]{burrows05,burrows06}. 

The observed visibilities are consistent with a partial resolution of
dust emission around HD69830.
Assuming that the stellar component has a visibility of 1, and that
the values of $F_\star$ and $F_{\rm{disc}}$ are accurately given by
a Kurucz model profile and the IRS spectra from \citet{beichman05},
the visibility ratios listed in Table \ref{tab:69830visdip} can be used
to estimate the disc location given certain assumptions about the morphology
of the emission.\footnote{A detailed description of how the models
  were compared to the observed visibility ratios and how errors were
  propagated through the analysis is given in Appendix A.}
For example, assuming that the disc emission has a Gaussian distribution
gives a best fit FWHM of 10mas. 
As many debris discs resolved to date have been found to have ring-like
structure \citep[see, e.g.][]{greaves,kalas05,schneider09}, we
also modelled the observed visibilities assuming the disc emission
to originate in symmetric, face-on rings with varying
radii and widths of $dr/r = 0.2, 1$ or 2 where $r$ is the radius of
the middle of the ring (see appendix A).
Figure \ref{fig:69830models} shows the results of a $\chi^2$-test of
the goodness-of-fit of each model suggesting a best-fit disc
radius of 10--25mas, depending on the width of the ring.
Both tested morphologies predict a best-fit radius of 0.1-0.3AU at 12.6pc
which would place the dust in the middle of the planetary system.
A stability analysis by \citet{lovis} used Monte-Carlo modelling to
determine locations where massless particles could have long-term
stable orbits in the presence of the 3 Neptune-mass planets,
finding 2 long-term stable zones at 0.3--0.5AU and $>$0.8AU.
Thus the visibilities suggest a best-fit location for the dust compatible
with the inner stable zone.

\begin{figure*}
\begin{minipage}{8cm}
\includegraphics[width=8cm]{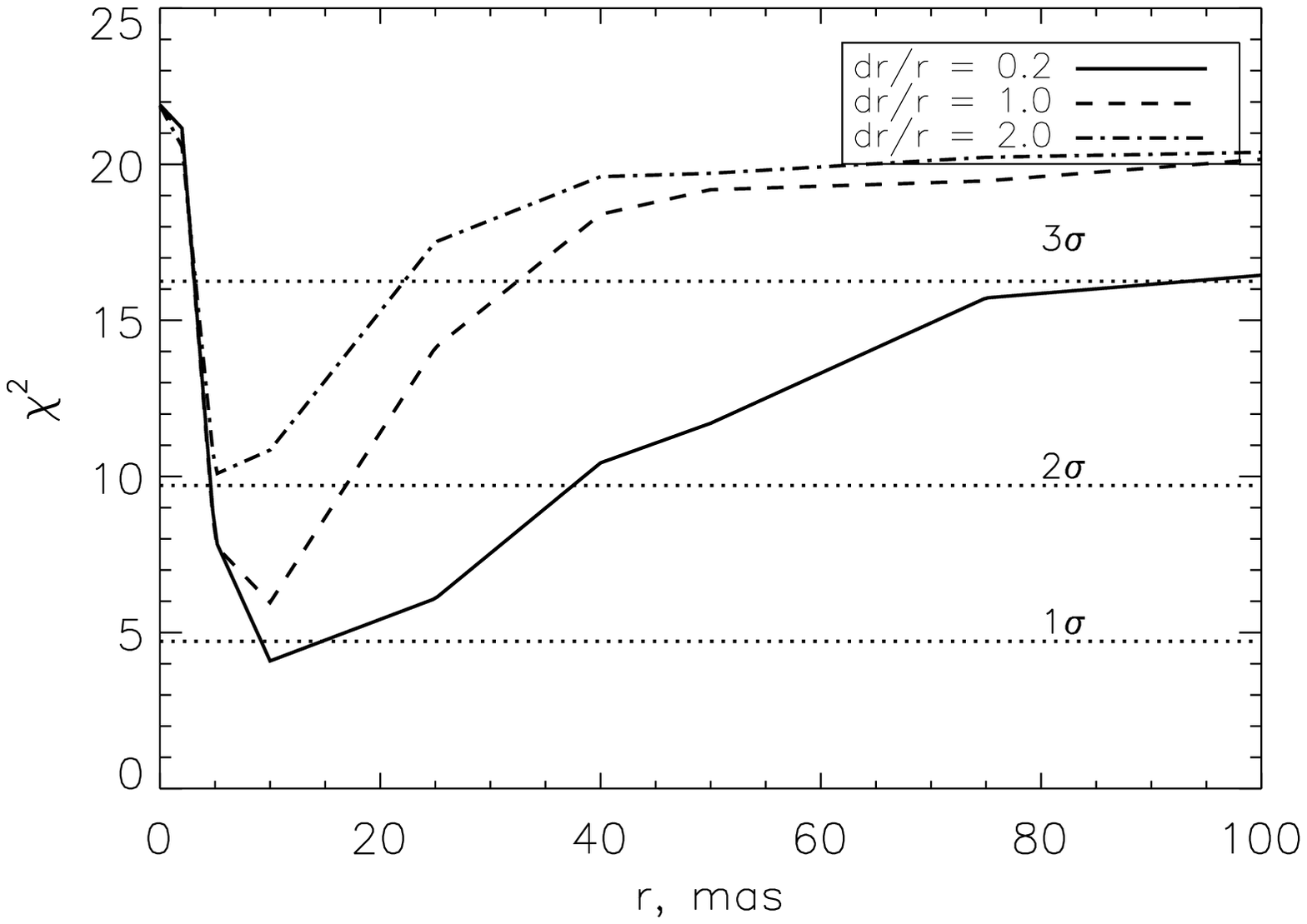}
\end{minipage}
\begin{minipage}{8cm}
\includegraphics[width=8cm]{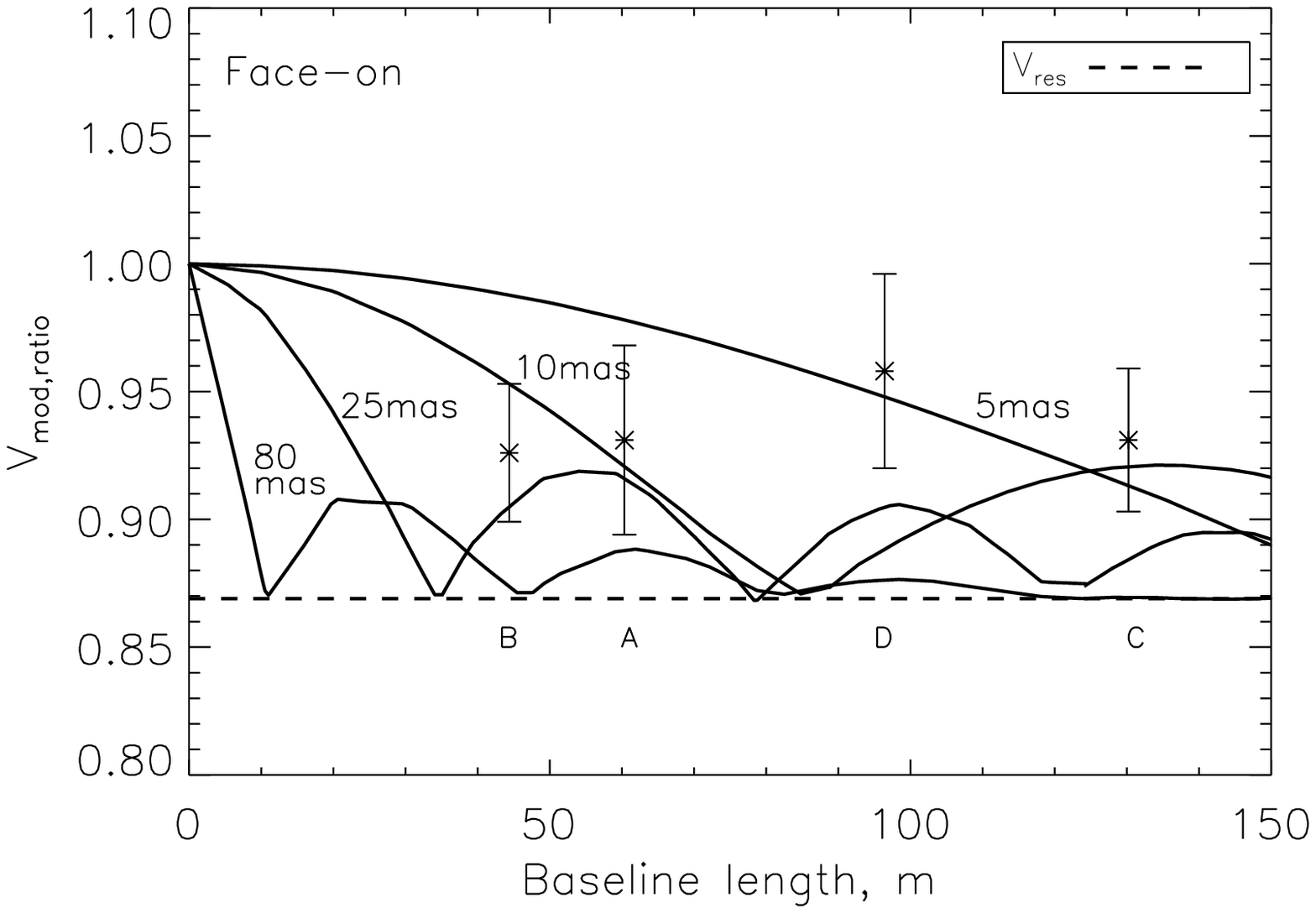}
\end{minipage}\\
\caption{\label{fig:69830models}
  The limits on the HD69830 disc from the MIDI observations.
  Left: The confidence limits found comparing the observed visibility
  ratios on all 4 baselines with those predicted by face-on ring-like models
  for the disc morphology (of radius $r$ and width $dr$).
  Lower values of $\chi^2$ represent a better fit to the data, and
  the horizontal dotted lines denote the level at which the models
  are consistent with the observations at the 1, 2 and 3$\sigma$
  levels as determined by the percentage points of the $\chi^2$
  distribution function with 4 degrees of freedom.
  The limits depend on the assumed width of the ring as indicated in the
  legend.
  Right: Predicted visibility ratios for narrow ($dr/r=0.2$) ring-like
  models of radii 5, 10, 25 and 80mas (solid lines).
  The ratio expected if the disc emission is completely resolved
  (i.e., the $V_{\rm{res}}$ prediction) is shown by a dashed line.
  Observed visibility ratios are shown with asterisks and 1$\sigma$
  error bars (Table \ref{tab:69830visdip}).
  The modest level of significance with which different models fit the 
  observations (seen in the left figure) is due to the complex
  visibility function for ring models, combined with sparse baseline
  coverage and large uncertainties.
  See Appendix A for details of modelling and error calculations.  }
\end{figure*}

However, uncertainties in the visibility ratios mean that the
observations are also consistent within the uncertainties
with a bigger radial location (see Figure \ref{fig:69830models}),
particularly if the possible slope on the visibilities is taken into account
(see section 3.1).  
Should the high levels of visibility seen at long wavelengths be
evidence of a residual slope, then the disc emission may be completely
resolved on these baselines (in which case the visibility would match
the $V_{\rm{res}}$ prediction).
For this to be the case we would require a Gaussian of FWHM $>$ 22mas
(0.28AU) on an average baseline of 83m, or of at least 40mas (0.5AU) to be
completely resolved on baseline B (44m).
These sizes are consistent both with the inner stable zone of
\citet{lovis}, and a larger radial location outside the orbits of the
planets.
A larger radial location is favoured from analysis of the IRS spectrum
which suggests the emission originates in disrupted P or D-type
asteroids at a distance of 0.9--1.1AU \citep{lisse07}.
If there is not a residual slope on the data, and the
results are truly indicative of a partial disc
resolution, the dust would then be constrained to lie within the
orbits of the planets challenging current SED models.

Regardless of the residual slope, the MIDI observations are inconsistent
with unresolved emission, and so limits can be placed on the minimum radius
of the dust emission. Using ring-like models we find a 3$\sigma$ limit
on the radius of $>$4mas (0.05AU), or from Gaussian models we find a
3$\sigma$ limit on the FWHM of $>$4mas.
The VISIR data (section 4) provides a complimentary upper limit on the
disc radius of $<$2.4AU.

\begin{figure}
\includegraphics[width=8cm]{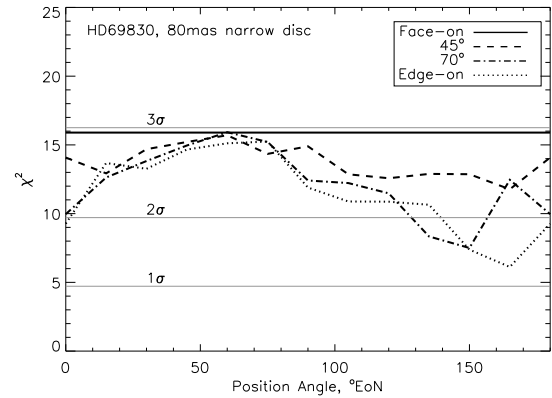}
\caption{\label{fig:698PAs} Comparison of inclined disc models to the
  observed visibility ratios of HD69830 at a fixed disc radius
  of 80mas.  Inclinations are measured from face-on = 0$^\circ$. }
\end{figure}

Adopting the radial size of 1AU from the SED modelling, we compared
the observations to model predictions for 1AU rings inclined both edge-on,
face-on and 45$^\circ$ to the line-of-sight and at different position angles
to assess whether the observations show evidence for an inclined disc
structure;
measurement of the disc inclination would remove ambiguity in the planet
masses assuming these were orbiting in the same plane.
A narrow edge-on disc lying at a position angle of
$\sim$165$^\circ$ provides the best fit to the four observed visibility
ratios ($\chi^2$ minimum of 1.3$\sigma$, Figure \ref{fig:698PAs};
other disc widths provide worse fits to the data, albeit still within
observational uncertainties). 
Inclined discs at other position angles away from $\sim$60$^\circ$
also provide better fits to the data than a face-on model, although
a narrow face-on disc at 80mas is consistent with the observed
visibilities at a level of 3$\sigma$.
Thus with the current data we cannot place significant limits on the
disc inclination.
The lack of significant differences in the visibility ratios on
different baselines also means there is no evidence for clumpy and
asymmetric structure, although the uncertainties in the data mean 
that such structure cannot be ruled out.

Taking the stability analysis of \citet{lovis} and the SED modelling of
\citet{lisse07} together with the MIDI and VISIR results presented here,
we consider the emission from this system to be most likely smooth and
symmetric at $\sim$1AU radius.

\subsection{$\eta$ Corvi}

\begin{figure*}
\begin{minipage}{8cm}
\includegraphics[width=8cm]{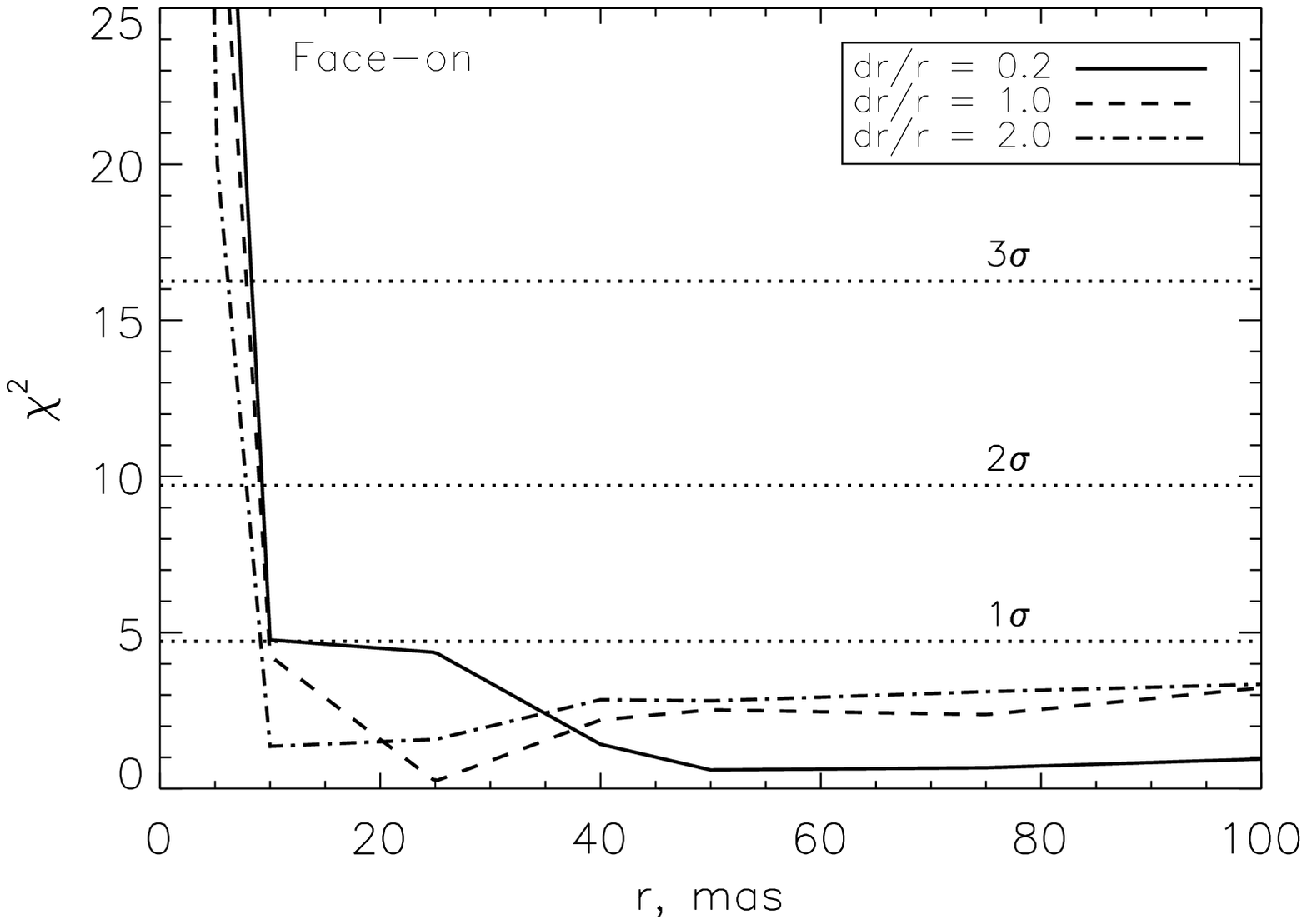}
\end{minipage}
\begin{minipage}{8cm}
\includegraphics[width=8cm]{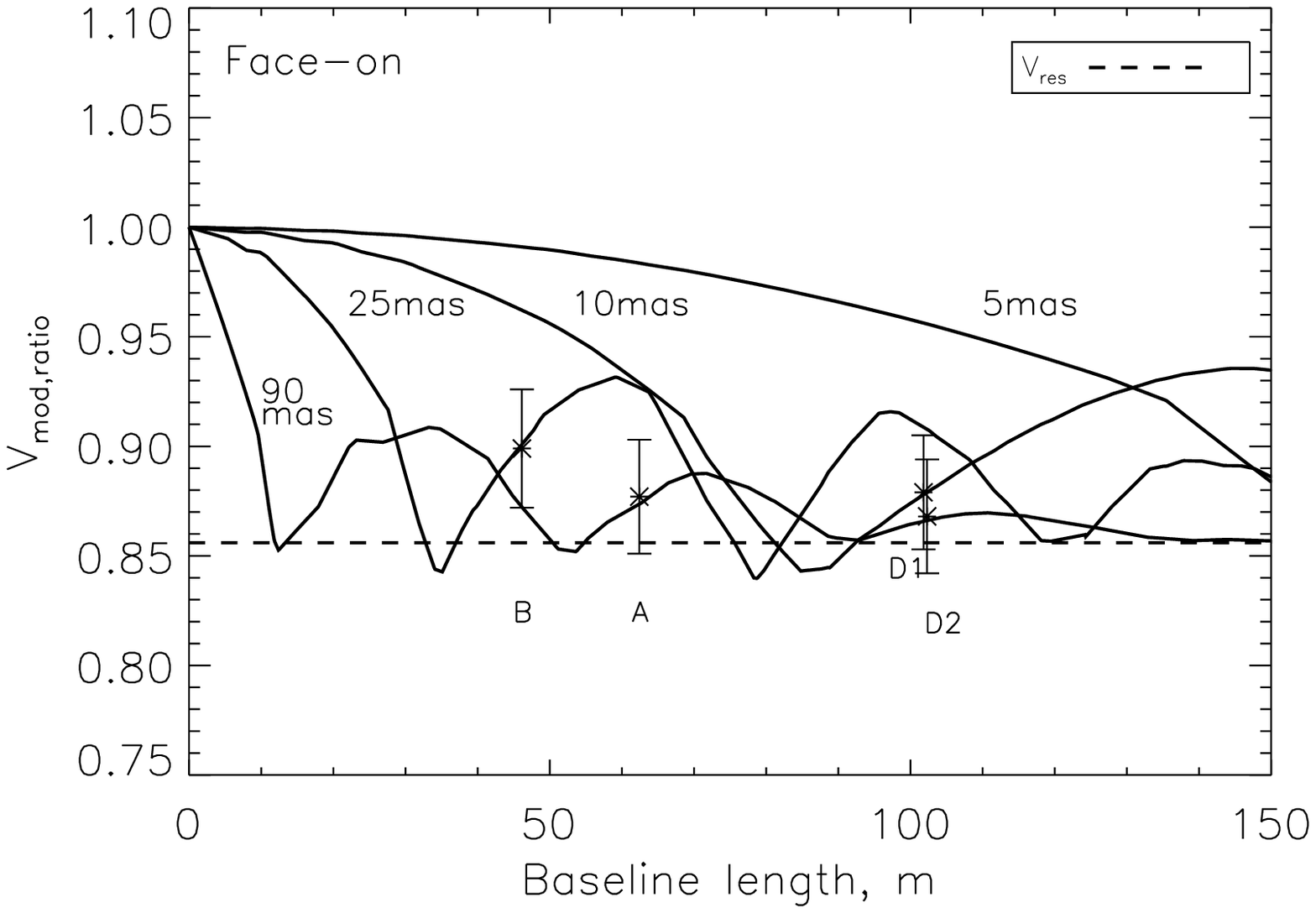}
\end{minipage}\\
\caption{\label{fig:etamodels}
  The limits on the $\eta$ Corvi warm dust from the MIDI observations.
  Left: The confidence limits found comparing the observed visibility
  ratios on all 4 baselines with those predicted by face-on ring-like models
  for the disc morphology (of radius $r$ and width $dr$).
  Lower values of $\chi^2$ represent a better fit to the data, and
  the horizontal dotted lines denote the level at which the models
  are consistent with the observations at the 1, 2 and 3$\sigma$
  levels as determined by the percentage points of the $\chi^2$
  distribution function with 4 degrees of freedom.
  The limits depend on the assumed width of the ring as indicated in the
  legend.
  Right: Predicted visibility ratios for narrow ($dr/r=0.2$) ring-like
  models of radii 5, 10, 25 and 90mas (solid lines).
  The ratio expected if the disc emission is completely resolved
  (i.e., the $V_{\rm{res}}$ prediction) is shown by a dashed line.
  Observed visibility ratios are shown with asterisks and 1$\sigma$
  error bars (Table \ref{tab:etadip}).
  The modest level of significance with which different models fit the 
  observations (seen in the left figure) is due to the complex
  visibility function for ring models, combined with sparse baseline
  coverage and large uncertainties.
  See Appendix A for details of modelling and error calculations.  }
\end{figure*}

The MIDI correlated fluxes are consistent with Rayleigh-Jeans
  emission across the MIDI range.  We see no evidence of the
  mid-infrared excess emission dominated by silicate emission in the
  correlated fluxes and conclude that the data are consistent with all
  of the excess being resolved on the observed baselines.
As for HD69830, we rule out a companion object as the source of the
resolved emission due to the silicate emission feature seen in the
IRS spectrum \citep{chen06} and the lack of a known binary companion.
Limits from radial velocity searches rule out a planet of mass
$>$0.4$M_{\rm{Jup}}$ with a period of 3 days or less, and a planet of mass
$>$2.1$M_{\rm{Jup}}$ with a period of 100 days or less (corresponding to
$<$0.46AU for $M_\star=1.5M_\odot$), at the 3$\sigma$ confidence level
\citep{lagrange08}. 

The observed visibilities are consistent with at least partial,
perhaps complete, resolution of the dust emission. 
The observed visibility ratios were modelled in a similar manner to
HD69830.
Assuming that the disc emission has a Gaussian distribution
results in a best fit FWHM of 27mas.
The comparison with ring-like models is shown in Figure \ref{fig:etamodels},
from which best-fit disc radii are 20--25mas for wider rings, or $>$50mas
for narrow face-on rings ($>$0.91AU).
A size of 20-50mas translates to a radius of 0.4-0.9AU at 18.2pc.
Since the MIDI observations are inconsistent with unresolved emission,
they can be used to derive a minimum size for the emitting region.
Assuming a Gaussian distribution for the disc emission limits the
FWHM to $>$14mas ($>$0.25AU; 3$\sigma$ limit from $\chi^2$
goodness-of-fit over all 4 baselines), whereas the ring-like models
constrains the radius to $>$9mas (0.16AU, $3\sigma$ limit).

There remains some uncertainty from SED modelling on the location of
the mid-infrared excess, with two models proposed that suggest that either 
the hot dust lies in a single ring at 1.7AU or in two radial locations
at 1.3 and 12AU \citep{chen06}.
\citet{smithhot} ruled out the 12AU component at the 2.6$\sigma$ level,
since this was not resolved in VISIR 18.7$\mu$m imaging, so that a single
ring at 1.7AU remains the best estimate of the radius of the hot dust 
from the SED, with unresolved 8m imaging constraining the location of
a single radius component to $<3$AU.
Thus, like HD69830, the best-fit radius is smaller than, but not
inconsistent with 
that predicted from SED modelling (90mas, 1.7AU, \citealt{smithhot});
the narrow ring models are most consistent with the SED models, however the
errors on the data and the complex visibility functions expected for
ring-like structure (Figure \ref{fig:etamodels}, right) mean that a large
range of ring models provide a good fit to the data within the
errors.
In contrast to HD69830 no planets are known to be orbiting interior
to the dust, although limits are not so stringent \citep{lagrange08}.

\begin{figure}
\includegraphics[width=8cm]{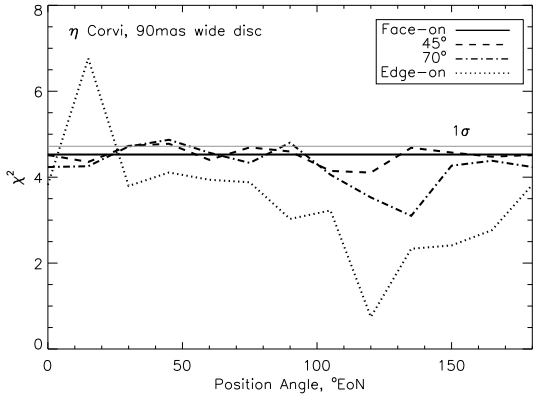}
\caption{\label{fig:etaPAs} Comparison of inclined disc models to the
  observed visibility ratios of $\eta$ Corvi at a fixed disc radius
  of 90mas.  Inclinations are measured from face-on = 0$^\circ$. }
\end{figure}

Adopting a radius of 1.7AU for the dust from SED modelling, we compared
the observations to model predictions for 1.7AU rings inclined both edge-on,
face-on and at 45$^\circ$ and 70$^{\circ}$ to the
line-of-sight and at different position angles to attempt to constrain
the inclination and position angle of such a disc. 
The best fitting disc geometry was found to be a broad edge-on
  disc at a position angle of $\sim$120$^\circ$ (Figure
  \ref{fig:etaPAs}).  For narrow discs all geometries are within
  1$\sigma$ of the best fitting model, and for broad discs only
  edge-on discs with a position angle $\sim$15$^\circ$ are outwith
  1$\sigma$ of the best fit.  Thus other disc geometries cannot be
  ruled out with the current data. 
However, it is notable that this best-fit position angle coincides
with that of a cold disc component (at 130$\pm$10$^\circ$) imaged
in the sub-mm and inferred to be in a 150$\pm$20AU ring inclined by 
45$\pm$25$^\circ$ to our line-of-sight \citep{wyatt05}.
A common position angle would suggest that both dust components, over
two orders of magnitude in radius, originate from planetesimals orbiting
in the same plane.

There is no evidence from this data for asymmetric of clumpy structure, since 
the visibility functions are consistent (within the errors) with the same 
level of resolution on all baselines (Figure \ref{fig:etaobsres}).
This level is consistent with the excess being completely resolved
on all baselines (i.e., it coincides with the $V_{\rm{res}}$ prediction).
We note that this does not mean that the emission comes from a region which 
appears the same size on all observed baselines, but that on each baseline 
the emission covers a large enough region to have been completely resolved.  

Taking the MIDI results and combining them with the results from 8m imaging 
in \citet{smithhot} and the IRS spectrum presented in \citet{chen06},
we consider the emission from this system to be most likely smooth and 
symmetric at 1.7AU radius, with tentative evidence for extension along a
position angle aligned with the cool outer disc.

\subsection{Transient sources of emission}

For both sources we infer the dust to lie at 1-1.7AU and so consider
that it must have a transient origin \citep{wyattsmith06}.
One possible source for the transient emission is a recent massive 
collision within a coincident planetesimal belt
\citep[a possibility proposed for BD+20307, see][and references
  therein]{zuckerman08}.  The emission from such an event
might be expected to be clumpy or asymmetric as the dust would start
off concentrated in a small spatial region (rather than dispersed in a
ring like an asteroid belt for example).
There is no evidence for such a structure from the MIDI observations.
Furthermore this scenario was considered as possible but unlikely
due to the expected low frequency of sufficiently massive
collisions and the lifetime of resulting dust particles from such events
\citep{wyattsmith06}. Thus it seems more likely that the parent
bodies of the dust seen at $\sim$1AU originated in a planetesimal belt
at larger distances where collisional timescales are much longer.
The mechanism transporting the dust into 1AU remains unknown, though one
possibility is a dynamical instability 
in a period analogous to the Late Heavy Bombardment (LHB) in the Solar
System \citep{gomes, booth}, the cause of which could be a
recent stellar flyby or the migration of giant planets in the system.

The known cool dust population toward $\eta$ Corvi \citep{wyatt05}
could represent the outer parent planetesimal belt suspected to be
feeding the hot dust, a suggestion supported by the tentative 
evidence from the MIDI observations for a position angle for the hot 
component (120$^\circ$) that is similar to that of the cool dust
(130$\pm$10$^\circ$).
However, the errors and sparse baseline coverage of the current data
do not allow significant limits to be placed at this stage on
the hot disc's inclination and position angle.
In contrast no cold dust population has yet been detected around HD69830.  
Indeed the MIPS photometry limits the excess emission at 70$\mu$m to 1$\pm$3 
mJy \citep{beichman05}, with SCUBA photometry limiting excess flux at
850$\mu$m to $<$7mJy \citep{sheret,matthews}, indicating that there is
little if any cold dust in the system.
One possible way around the conclusion that the dust has to be transient
was suggested by \citet{payne}, who considered the dynamical 
evolution of the HD69830 planetesimal disc during the growth and 
migration its 3 planets \citep{alibert}.
In their model significant planetesimal mass remained outside 1AU
on high eccentricity orbits, and they suggested that the increased
collisional lifetime of highly eccentric orbits could allow significant
mass to persist even to 2Gyr, a possibility which is currently
being investigated.


\section{Conclusions}

In this paper we present MIDI observations of HD69830 and
$\eta$ Corvi, which for the first time resolve the hot dust
emission in these systems, and setting lower limits on the
radial location of the emission of $>$0.05AU and $>$0.16AU,
respectively.
We also present new VISIR observations of HD69830 which show
no evidence for resolved emission, thus constraining the radial location
of the emission to $<$2.4AU.
Similar 8m observations of $\eta$ Corvi were already presented in
\citet{smithhot} showing this emission originates $<$3AU.
These mid-infrared observations thus place the dust squarely
at a location consistent with that predicted from SED modelling at
1.0 and 1.7AU for HD69830 and $\eta$ Corvi respectively.
At such a location the emission to both sources is expected to be
transient.

There is no evidence from the MIDI data that the emission is
anything but smooth and symmetric.
However, assuming a 1.7AU radius for the emission toward $\eta$ Corvi
the data marginally favour a more edge-on disc lying at a position angle
consistent with that of the known 150AU cool disc at 130$\pm$10$^\circ$.
Such a correlation would further support the suggestion that the
cool outer disc is feeding the inner disc.
The MIDI data also suggest that the HD69830 emission is partially,
rather than completely, resolved, placing it at 0.1--0.3AU.
Observational uncertainties, including a possible residual slope,
mean that this interpretation is not favoured at present given
the SED constraints, but if confirmed would present a challenge
to the current modelling of the IRS spectra of this source.  

These observations highlight the viability of exploration of
dust in terrestrial planet regions using mid-infrared interferometry.
Future MIDI observations of HD69830 and $\eta$ Corvi that extend
the baseline coverage would allow more stringent constraints to
be placed on the geometry of these debris discs, including measuring
the disc inclination for HD69830 (thus constraining the masses of
planet found by radial velocity studies), the disc position angle 
for $\eta$ Corvi (for comparison with that of the outer cool disc),
and searching for the clumpy or asymmetric structure that
might be expected from a recent massive collision.
Such observations thus have significant potential to constrain
the different theories for the origin of these unusual hot dust
populations.


\begin{acknowledgements}
RS is grateful for the support of a Royal Commission for the
Exhibition of 1851 Fellowship.  
The authors wish to extend their thanks to Christine Chen for providing
IRS spectra of $\eta$ Corvi.  We also wish to
thank Chas Beichman for providing the IRS spectra of HD69830.  
Based on observations made with ESO Telescopes at the Paranal
Observatory under programme IDs 078.D-0808 and 079.C-2059.    
\end{acknowledgements}

\bibliographystyle{aa}  
\bibliography{/home/rachel/Work/PAPERs/thesis} 

\begin{appendix}
\label{s:app}

\section{Modelling the MIDI observations}

In this appendix we outline the method used to compare models for
geometry of the excess emission to the MIDI data.  We consider
symmetric Gaussians of varying FWHM and ring models. The ring models 
are face-on axisymmetric discs with a fixed width and varying radius.
The disc is assumed to be centred on the point-like star and has
uniform surface brightness.  The tested widths are $dr/r  = 0.2, 1$
or 2, with different disc radii, parameterised by $r$, the mid-point
of the 
disc (so disc has inner radius $r-dr/2$, outer radius $r+dr/2$) with
$r$ between 5--200 mas.  Due to the large range of possible
non-axisymmetric disc models (for example discs with brightness
asymmetries or clumps, or those inclined to the line of sight at all
position angles) we consider only a restricted range of discs outside
these face-on models.  These were edge-on discs and those inclined at
45 and 70$^\circ$ to the line-of-sight.  These discs were assumed to have a
fixed radius (80mas for HD69830, 90mas for $\eta$ Corvi; sizes from
SED fitting) and had position angles on sky of 0--180$^\circ$ East of
North in steps of 15$^\circ$.  

The excess region (Gaussian or ring) visibilities were determined
using the \emph{Interferometric  Visibility Calculator} available at
http://www.mporzio.astro.it/\%7Elicausi/IVC/. Determination of the
total model visibility then follows from 
equation \ref{eq:calcvis}, assuming the visibility of the star is 1 at
all wavelengths.  We determined the visibility for the models across
the wavelength ranges used in the characterisation of the dips
(i.e. 8--9$\mu$m and 10--11.5$\mu$m, Tables \ref{tab:69830visdip} and
\ref{tab:etadip}).  The models assume flux ratios for the star and
disc components are accurately given by the IRS photometry and the
Kurucz model profiles.  

\subsection{Applying a goodness-of-fit test}

As the absolute levels of visibility are poorly constrained but the
ratio of visibilities at different wavelengths is constrained much
better (see 
sections 2.4, 3.1 and 3.2), we shall use comparison between the
observed and model visibility ratios to determine the goodness-of-fit
of different models.  The sources of error for this comparison must be
carefully considered to allow an accurate goodness-of-fit
calculation.

In this paper we have used the IRS photometry to calculate
visibilities.  Following equations \ref{eq:calcorr} and
\ref{eq:calvis}, each observed visibility is calculated as  
\begin{equation}\label{eq:myviscalc}
V = I_{\rm{corr,tar}}/I_{\rm{corr,cal}} \times
F_{\rm{tot,cal}}/F_{\rm{tot,tar}} \times V_{\rm{cal}}.
\end{equation}
When we consider the ratio of visibilities, we divide the two
visibilities as averaged over the appropriate wavelength ranges
\begin{equation}\label{eq:visobsrat}
V_{\rm{obs,ratio}} = \frac{I_{\rm{corr,tar,10}}}{I_{\rm{corr,tar,8}}}
\times 
\frac{F_{\rm{tot,cal,10}}}{F_{\rm{tot,cal,8}}} \times 
\frac{I_{\rm{corr,cal,8}}}{I_{\rm{corr,cal,10}}} \times
\frac{F_{\rm{tot,tar,8}}}{F_{\rm{tot,tar,10}}} \times 
\frac{V_{\rm{cal,10}}}{V_{\rm{cal,8}}}
\end{equation}
where the subscript 10 refers to the value of this term averaged over
the wavelength range 10--11.5$\mu$m, and the subscript 8 refers to
averaging over the range 8--9$\mu$m.  We now consider the errors from
each term in the above equation. 

The first term in equation \ref{eq:visobsrat}, 
$\frac{I_{\rm{corr,tar,10}}}{I_{\rm{corr,tar,8}}}$, is
from the correlated fluxes as measured in the fringe tracking. 
Errors in each wavelength range are given
by variation in the five sub-integrations of the fringe tracking (as
given in Tables \ref{tab:69830visdip} and \ref{tab:etadip}) and
are propagated as $\sigma_{I, \rm{ratio}}/I_{\rm{ratio}} =
\sqrt{(\sigma_{I,8}/I_8)^2 + (\sigma_{I,10}/I_{10})^2}.$  The second
term, $\frac{F_{\rm{tot,cal,10}}}{F_{\rm{tot,cal,8}}}$, is given by
Kurucz model profile fits to the standard stars.  As the slope is
considered to be well modelled by such profiles in the mid-infrared
and 
there are no spectral features expected in this range for the
standards we consider errors on this term to be negligible.  The third
term, $\frac{I_{\rm{corr,cal,8}}}{I_{\rm{corr,cal,10}}}$, comes from
the standard star correlated flux measurements.  As shown in section
2.4 this ratio should always be 1 but has an error of 3.7\%.  In those
instances in which an average calibration from two standard star
observations was used this error can be reduced by a factor of
$\sqrt{2}$. The term
$\frac{F_{\rm{tot,tar,8}}}{F_{\rm{tot,tar,10}}}$ is given by the IRS
spectrum of each target.  For this we must consider how well the IRS
spectra captures the shape of a spectrum.  \citet{beichman06} showed
that  for Sun-like stars with no excess emission deviations from a
photospheric slope were 1.3\% in the range 8.5--13$\mu$m 
after removing the effects of absolute calibration error (by scaling
to the photospheric model at 7--8$\mu$m). The stars included in this
study are of the same or 
similar spectral type to HD69830 (which was in fact included in the
sample). $\eta$ Corvi was just excluded by spectral type;
the paper considered stars in the range F5--K5, but we do not consider
an F2 star likely to be very different to this sample.  We therefore
use the error of 1.3\% as the error on the total flux ratio for the
science targets. The term $\frac{V_{\rm{cal,10}}}{V_{\rm{cal,8}}}$ is
determined by the assumed size of the standard star used for
calibration and in these calculations is assumed to be perfect.  In
fact as can be seen in Table \ref{tab:sources} there is an uncertainty
on the size of the standard stars.  This uncertainty is very small
however (typically 0.01mas) and potential error on the calibrator
visibility ratio is thus very small ($<$0.01\%).  We therefore neglect
this uncertainty in our error calculations.

To calculate the same ratio for each model we must use the following
equation (c.f. equation \ref{eq:calcvis}):
\begin{eqnarray}\label{eq:vismodrat}
V_{\rm{mod,ratio}}  & =  &
\frac{(F_{\star\rm{,tar,10}}/F_{\rm{tot,tar,10}}) + 
(F_{\rm{disc,tar,10}}/F_{\rm{tot,tar,10}})V_{\rm{disc,tar,10}}} 
{(F_{\star\rm{,tar,8}}/F_{\rm{tot,tar,8}}) +
(F_{\rm{disc,tar,8}}/F_{\rm{tot,tar,8}})V_{\rm{disc,tar,8}}} \nonumber
\\ & = & \frac{F_{\star\rm{,tar,10}}}{F_{\star\rm{,tar,8}}} \times
\frac{F_{\rm{tot,tar,8}}}{F_{\rm{tot,tar,10}}} \times
\frac{ 1 + V_{\rm{disc,tar,10}}\left(
  \frac{F_{\rm{tot,tar,10}}}{F_{\star\rm{,tar,10}}} - 1 \right)}
   {1 + V_{\rm{disc,tar,8}}\left(
  \frac{F_{\rm{tot,tar,8}}}{F_{\star\rm{,tar,8}}} - 1 \right)} 
\end{eqnarray}
where it has been assumed the $V_{\star}$ is 1 in both wavelength
ranges (see section 3).  
The errors on each term for the model ratio also require careful
consideration. The first term in equation \ref{eq:vismodrat}, 
$\frac{F_{\star\rm{,tar,10}}}{F_{\star\rm{,tar,8}}}$, comes from Kurucz
model profiles, 
and again as the stellar emission in this range can be considered to
have 
a well-defined slope the error on this term is negligible. The
second term, $\frac{F_{\rm{tot,tar,8}}}{F_{\rm{tot,tar,10}}}$, again
comes from 
the IRS spectrum of this source, and thus we again consider the errors
on this ratio to be 1.3\% (see above).  

The errors on the final term in equation \ref{eq:vismodrat} are more
complex.  We consider that for each model the visibility
$V_{\rm{disc,tar}}$ is perfectly known.  However errors on
$F_{\rm{tot,tar}}/F_{\star\rm{,tar}}$ in both wavelength ranges come
from a variety of sources, some of which are systematic across the
different wavelength ranges considered.  In order to determine the
error on this factor we used a Monte-Carlo method.  For each model
(each with a fixed $V_{\rm{disc,tar,8}}$ and $V_{\rm{disc,tar,10}}$
for each baseline) the value of this final factor was calculated over
1000 iterations, with the values of $F_{\rm{tot,tar}}$ and
$F_{\star\rm{,tar}}$ chosen as follows.   The values of
$F_{\star\rm{,tar,8}}$ and $F_{\star\rm{,tar,10}}$ were determined by
using the Kurucz model profile for the correct spectral type (see
Table \ref{tab:sources}).  Scaling was determined by taking the
$\chi^2$ minimisation over a range of scalings to find the best fit
to the B- and V-band magnitudes of the stars as listed in the Hipparcos
catalogue and J-, H- and K-band magnitudes from the 2MASS catalogue.
The percentage points of the $\chi^2$ distribution were used to
determine a 1$\sigma$ limit on this scaling.  The scaling used in the
Monte-Carlo testing was then taken from a normal distribution with a
mean given by the best fitting scaling and standard deviation at the
level of the 1$\sigma$ error.  The $F_{\rm{tot,tar,8}}$ and $F_{\rm{tot,tar,10}}$ 
values were determined by averaging over the appropriate wavelength
ranges after the flux in each spectral channel was randomly sampled
from a normal distribution with a mean of the IRS total flux in
the channel and standard deviation the error in the channel.  To
approximate the effect of calibration error the values of
$F_{\rm{tot,tar,8}}$ and $F_{\rm{tot,tar,10}}$ were then multiplied by a factor
randomly sampled from a normal distribution with mean 1 and standard
deviation 0.05 (calibration uncertainty for IRS typically 5\% at short
wavelengths; \citealt{FEPSexplan}); the same ``calibration value'' was
used for $F_{\rm{tot,tar,8}}$ and $F_{\rm{tot,tar,10}}$.  The mean value of this
final factor in equation \ref{eq:vismodrat} was checked against the
its value assuming no errors on $F_{\rm{tot,tar}}$ and $F_{\star\rm{,tar}}$ and
found to be the same.  The error term on this final factor was
then taken from the standard deviation over the 1000 iterations, and
was found to be typically at the 1\% level for both sources averaged
over all disc model and baseline combinations.

To test how well each model reproduces the observed visibility ratios
we use the $\chi^2$ goodness-of-fit test.  Typically this takes the
form $\Sigma_{n=1}^{N} (D-M)^2/\sigma^2$ where N is the number of data
points, $D$ is some observed data with associated error $\sigma$ and
$M$ is a model with no error.  As both $V_{\rm{obs,ratio}}$ and
$V_{\rm{mod,ratio}}$ have errors associated with them, we use the
ratio of the observed to model visibility ratios and compare them to
1 (as if the model is a good fit to the data then
$V_{\rm{mod,ratio}} \approx V_{\rm{obs,ratio}}$).  The errors from
both visibility ratios must be included in the value of $\sigma$.  Our
$\chi^2$ calculation then becomes
\begin{equation}\label{eq:chisq}
\chi^2 = \sum_{i=1}^{4}
\left(\frac{V_{\rm{obs,ratio}}(i)/V_{\rm{mod,ratio}}(i)-1}{\sigma_i}\right)^2 
\end{equation} 
where $i$ labels the observations of each source, and $\sigma_i$ is
the error on $V_{\rm{obs,ratio}}/V_{\rm{mod,ratio}}$.  Consideration
of equations \ref{eq:visobsrat} and \ref{eq:vismodrat} shows that the
$F_{\rm{tot,tar}}$ term cancels when taking the ratio of the observed and
model visibility calculations, and thus the 1.3\% error term does not
explicitly appear in our $\chi^2$ calculations. Errors on the IRS
photometry do appear implicitly in the Monte-Carlo calculation of the
error on the final term in equation \ref{eq:vismodrat} through the
calibration error and the statistical errors.
The final value of $\sigma$ comes 
from addition in quadrature of all the remaining error terms outlined
in the above paragraphs. We use the percentage points of the $\chi^2$
distribution to determine the values of $r$ that provide a reasonable
fit to the observed data for both sources.  The results of this
modelling are discussed in sections 5.1 and 5.2 and shown in Figures
\ref{fig:69830models} and \ref{fig:etamodels}.

\end{appendix}

\end{document}